\documentclass[%
 aip,
 amsmath,amssymb,
 reprint,%
]{revtex4-1}

\usepackage{graphicx}
\usepackage{dcolumn}
\usepackage{bm}
\usepackage{subfigure} 
\usepackage[utf8]{inputenc}
\usepackage[T1]{fontenc}
\usepackage{mathptmx}
\usepackage{etoolbox}

\usepackage{color,soul}
\usepackage{ulem}

\makeatletter\def\@email#1#2{%
 \endgroup
 \patchcmd{\titleblock@produce}
  {\frontmatter@RRAPformat}
  {\frontmatter@RRAPformat{\produce@RRAP{*#1\href{mailto:#2}{#2}}}\frontmatter@RRAPformat}
  {}{}
}%
\makeatother
\begin{document}

\preprint{AIP/123-QED}

\title[double conical capillary]{
Thermodynamics of imbibition in capillaries of double conical structures-Hourglass, diamond, and sawtooth shaped capillaries-
}
\author{Masao Iwamatsu}
 \email{iwamatm@tcu.ac.jp}
\affiliation{ 
Tokyo City University, Setagaya-ku, Tokyo 158-8557, Japan
}%


\date{\today}

\begin{abstract}
Thermodynamics of imbibition (intrusion and extrusion) in capillaries of double conical structures is theoretically studied using the classical capillary model. By extending the knowledge of the thermodynamics of a single conical capillary, not only the nature of spontaneous imbibition but that of forced imbibition under applied external pressure are clarified. Spontaneous imbibition in capillaries of double conical structure can be predicted from the Laplace pressure in a single conical capillary. To understand the forced imbibition process, the free energy landscape along the imbibition pathway is calculated. This landscape shows either a maximum or a minimum.  The former acts as the energy barrier and the latter acts as the trap for the liquid-vapor meniscus so that the imbibition process can be either abrupt with a pressure hysteresis or gradual and continuous. The landscape also predicts a completely filled, a half-filled and a completely empty state as the thermodynamically stable state. Furthermore, it also predicts a completely filled and a half-filled state of metastable liquid which can be prepared by the combination of the intrusion and the extrusion process. Our study could be useful for understanding various natural fluidic systems and for designing functional fluidic devices such as a diode, a switch etc. 
\end{abstract}

\maketitle

%

\section{\label{sec:sec1}Introduction}

Imbibition (intrusion and extrusion) of liquid in microscale and nanoscale capillaries is one of the most fundamental problems of thermodynamics of liquid in confined space not only in various field of natural science~\cite{Prakash2008,Kim2012,Tinti2017,Donne2022,Cai2022,Tinti2023} but also in various engineering problems in micro- and nano-scale~\cite{Squires2005,Bocquet2010,Haywood2014,Fraux2017,Kavokine2021,Wang2022,Robin2023}. Recently, there have been growing interests in addressing the problem of imbibition in asymmetric capillaries with geometrical gradient~\cite{Cai2022} because it is relevant to the engineering~\cite{Comanns2015,Li2017a,Buchberger2018} of various micro- and nano-fluidics functional devices.

\begin{figure}[htbp]
\begin{center}
\includegraphics[width=0.8\linewidth]{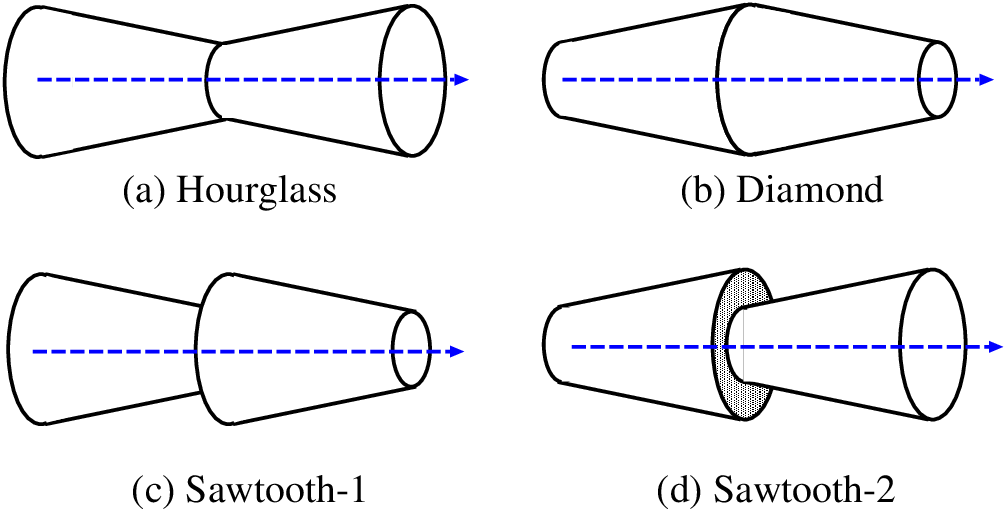}
\end{center}
\caption{
Four double conical capillaries consisting of two truncated conical capillaries of identical shape.  Arrows indicate the direction of intrusion of liquid.
}
\label{fig:D1}
\end{figure}

Among various asymmetric capillaries, conical and double-conical capillaries~\cite{Chuang2023} illustrated in Fig.~\ref{fig:D1} are the basic element of various natural as well as artificial systems. In particular, truncated conical capillaries have been extensively studied as the simplest model to study the effect of geometrical gradient and, in particular, as the model of imbibition into
porous substrates~\cite{Cai2022}.  Also, they have been studied for their potential applications as micro- and nano-fluidic devices~\cite{Wang2022,Buchberger2018,Cervera2006,Zhang2017,Singh2020,Iwamatsu2022,Xu2023,Leivas2023} such as liquid diode~\cite{Wang2022,Buchberger2018,Singh2020,Iwamatsu2022}, ionic current rectifier~\cite{Cervera2006}, pump~\cite{Zhang2017}, Janus paper~\cite{Xu2023}, and water harvesting~\cite{Leivas2023}. Carbon nanocone seems the most promising candidate of conical nano capillaries~\cite{Li2018,Leivas2023}.

Those capillaries in Fig.~\ref{fig:D1} with double conical structures which consist of a converging and a diverging conical capillary also attract intensive attentions recently. For example, {\it converging-diverging} hourglass shaped capillaries illustrated in Fig.~\ref{fig:D1}(a) have been studied as the simplest model of biological aquaporin~\cite{Gravelle2013} and its biomimetic artificial devices for filters~\cite{Balannec2018}, pumps~\cite{Antunes2022}, gates~\cite{Trick2014} and rectifiers~\cite{Li2019}. To fabricate hourglass shaped capillaries, mechanical deformation of carbon nanotube has been considered~\cite{He2014,Cao2019}. In addition to the flow physics in converging-diverging capillaries~\cite{Goli2022}, the flow physics in similar {\it diverging-converging} diamond shaped capillaries (Fig.~\ref{fig:D1}(b)) attract some attentions as a model fluidic device~\cite{Goli2019} and as a theoretical conceptual tool~\cite{Schimmele2007}.

In addition to those converging-diverging hourglass shaped and diverging-converging diamond shaped structure, {\it converging-converging} sawtooth shaped structure (Sawtooth-1 in Fig.~\ref{fig:D1}(c) and {\it diverging-diverging} sawtooth shaped structure (sawtooth-2 in Fig.~\ref{fig:D1}(d) have attracted attentions~\cite{Prakash2008,Kim2012,Comanns2015, Li2017a,Buchberger2018} for their ratchet-like structure, which is expected to realize unidirectional transport.

Although a large amount of literature on double conical capillaries has already been accumulated~\cite{Squires2005,Bocquet2010,Haywood2014,Kavokine2021,Cai2022,Wang2022,Robin2023}, most of the theoretical works studied transport properties numerically using macroscopic fluid dynamic equations~\cite{Cervera2006,Zhang2017,Chuang2023,Goli2019,Singh2020,Goli2022,Antunes2022} or atomic molecular dynamic simulations~\cite{Li2018,Leivas2023,Trick2014,Li2019} which are limited to sub nano-scale.  Relatively few studies based on thermodynamics have been conducted to understand the quasi-static imbibition processes~\cite{Lefevre2004,Remsing2015,Kaufman2017,Panter2020,Iwamatsu2020,Iwamatsu2022,Donne2022}. In particular, pressure-controlled intrusion and extrusion, or infiltration and defiltration, and, furthermore, infiltration pressure~\cite{Fraux2017,Liu2009,Goldsmith2009,Mo2015,Cheon2023} must be important not only to understand macroscopic natural and geographic problems~\cite{Cai2022} but also to design micro- and nano-fluidic devices.

In our previous paper~\cite{Iwamatsu2022}, we used thermodynamic approach and derived an analytical formula for the modified Laplace pressure and the free energy landscape of imbibition in a diverging and a converging conical capillary. Then, we can determine the criterion for the appearance of diode-like character (one-way transport) in a single conical capillary.  In contrast, many researchers~\cite{Urteaga2013,Berli2014,Gorce2016,Singh2020} used hydrodynamic approach~\cite{Reyssat2008} and identified diode-like character in conical capillaries from the time scale of flow. However, such hydrodynamic studies can be meaningful only when the spontaneous imbibition (intrusion) without external applied pressure is realized and the steady flow is established.

Furthermore, we found that the free energy landscape shows either a maximum which acts as the barrier or a minimum which acts as the trap so that the imbibition can be either an abrupt transition with a hysteresis or a gradual continuous transition~\cite{Iwamatsu2020,Iwamatsu2022}. This free energy maximum originates from the conical geometry and is nothing related to the nucleation barrier of capillary condensation~\cite{Lefevre2004,Remsing2015,Tinti2017,Tinti2023} because the free energy landscape is evaluated by assuming a continuous intrusion of liquid from one end of the capillary~\cite{Iwamatsu2020,Iwamatsu2022}.

In this paper, we enlarge our previous studies of a single conical capillary~\cite{Iwamatsu2020,Iwamatsu2022}, and consider the thermodynamics of imbibition in doble conical capillaries. Here, the terminology "imbibition" collectively used to mean "intrusion" and "extrusion" of liquid. To this end, we consider not only the Laplace pressure\cite{Lefevre2004,Panter2020,Iwamatsu2022} which is the main driving force of capillary flow~\cite{Washburn1921,Landau1987} but also the free energy landscape along the pathway of imbibition~\cite{Tinti2017,Iwamatsu2020,Iwamatsu2022} under the applied external (infiltration) pressure.  Our results will be useful to consider the possibility of various double conical capillaries as the functional liquid devices by pressure-controlled imbibition under the action of the infiltration pressure~\cite{Liu2009,Goldsmith2009,Mo2015,Fraux2017,Cheon2023}.

\section{\label{sec:sec2}Imbibition in a converging and a diverging single conical capillary}
\subsection{\label{sec:sec2.1}Morphological thermodynamics of imbibition}

In this section we reconsider and enlarge our previous studies~\cite{Iwamatsu2020,Iwamatsu2022} of the classical capillary model of imbibition (intrusion and extrusion) in a single conical capillary. Though the classical capillary model, which is the simplest case of morphological thermodynamic approach~\cite{Konig2004,Roth2006}, is macroscopic, it is believed to be valid down to the nanoscale~\cite{Tinti2023} and would give useful information even to the micro and the nano scale phenomena.  In this classical model, the surface free energy $F$ comprises the free energy of the free liquid-vapor surface energy $F_{\rm lv}=\gamma_{\rm lv}S_{\rm lv}$ and that of the liquid-solid surface energy of the capillary wall $F_{\rm sl}=\gamma_{\rm lv}\cos\theta_{\rm Y}S_{\rm sl}$ wetted by the liquid. The total surface free energy is given by
\begin{equation}
F=F_{\rm lv}-F_{\rm sl}=\gamma_{\rm lv}S_{\rm lv}-\gamma_{\rm lv}\cos\theta_{\rm Y}S_{\rm sl},
\label{eq:D1}
\end{equation}
where $\gamma_{\rm lv}$ and $S_{\rm lv}$ represent the liquid-vapor surface tension and the surface area, respectively, and $S_{\rm sl}$ is the solid-liquid (wet) surface area (see Tab.~\ref{tab:D1} for the complete list of symbols and their description.). The angle $\theta_{\rm Y}$ is Young's contact angle defined by Young's equation, which is expressed as:
\begin{equation}
\cos\theta_{\rm Y}=\frac{\gamma_{\rm sv}-\gamma_{\rm sl}}{\gamma_{\rm lv}},
\label{eq:D2}
\end{equation}
where $\gamma_{\rm sv}$ and $\gamma_{\rm sl}$ represent the solid-vapor and the solid-liquid surface tensions, respectively. This Young's contact angle characterizes the wettability of capillary wall. Here, we neglect the effect of gravity since we consider capillaries whose diameters are smaller than capillary length. Also, we neglect the contribution of the line tension, which could play some role in nano scale.

\begin{table}
\caption{
Symbols used
}
\label{tab:D1}
\begin{tabular}{lc}\hline\hline
Symbol & Description \\
\hline
$\gamma_{\rm lv}$ & Liquid-vapor surface tension \\
$\gamma_{\rm sl}$ & Solid-liquid surface tension \\
$\gamma_{\rm sv}$ & Solid-vapor surface tension \\
$S_{\rm lv}$ & Liquid-vapor surface area \\
$S_{\rm sl}$ & Solid-liquid (wet) surface area \\
$F$ & Total surface free energy \\
$F_{\rm lv}$ &   Liquid-vapor surface free energy \\
$F_{\rm sl}$ &   Solid-liquid surface free energy \\
$\theta_{\rm Y}$ & Young's contact angle \\
$\psi$ & Half-opening angle of spherical meniscus \\
$v(\psi)$ & Small volume correction of spherical cap \\
$p_{\rm L}(z)$ & Liquid (capillary) pressure \\
$z$ & meniscus position \\
$\phi$ & Tilt angle of conical wall \\
$H$ & Length of conical capillary \\
$R_{i}(z)$ & Radius of converging ($i=$C) and diverging ($i=$D) capillary \\
$\eta_{i}$ & Aspect ratio \\
$p_{{\rm L}(i)}(z)$ &  Liquid (capillary) pressure\\
$\Pi_{i}\left(\theta_{\rm Y},\phi\right)$ & Scaled pressure \\
$p_{\rm ext}$ & External pressure \\
$p_{i}(z)$ & Liquid (capillary) pressure of forced imbibition \\
$p_{{\rm c}(i)}$ & Critical external pressure \\
$\Omega_{i}$ & Grand free energy landscape \\
$\omega\left(\tilde{z}\right)$ & Non-dimensional grand free energy landscape \\
$\tilde{\omega}\left(\tilde{z}\right)$ & Non-dimensional grand free energy landscape \\
$\tilde{z}$ & Non-dimensional meniscus position \\
$\tilde{p}$ & Non-dimensional external pressure \\
$\alpha_{i}$ & Parameter which characterizes conical shape \\
$p_{{\rm e}(i)}$ & Characteristic external pressure \\
$p_{{\rm s}(i)}$ & Characteristic external pressure \\
$\tilde{z}_{\rm ex}$ & Position of extremum of free energy landscape \\
$\omega_{\rm ex}$ & Free energy at the extremum of free energy landscape \\
$\tilde{\omega}_{\rm CD}(\tilde{z})$ & Free energy landscape of converging-diverging hourglass \\
$\tilde{\omega}_{\rm DC}(\tilde{z})$ & Free energy landscape of diverging-converging diamond \\
$\tilde{\omega}_{\rm CC}(\tilde{z})$ & Free energy landscape of converging-converging sawtooth-1 \\
$\tilde{\omega}_{\rm DD}(\tilde{z})$ & Free energy landscape of diverging-diverging sawtooth-2 \\
\hline
\end{tabular}
\end{table}

Even though we will not consider the nucleation of liquid droplets or vapor bubbles in the middle of capillary, we will use the terminology "vapor" instead of "gas" throughout this paper.  In fact, our model system can be applicable to any binary system of immiscible fluids including non-volatile liquid and gas systems.

\begin{figure}[htbp]
\begin{center}
\includegraphics[width=0.4\linewidth]{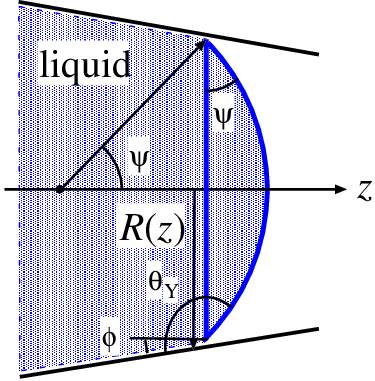}
\end{center}
\caption{
Spherical meniscus of liquid in a capillary axially symmetric around $z$ axis with a varying radius $R(z)$. The half-opening angle $\psi$ of the spherical interface is related to the tilt angle $\phi$ and Young's contact angle $\theta_{\rm Y}$ of the capillary wall. Here, we show the convex meniscus with the contact angle $\theta_{\rm Y}$ larger than $90^{\circ}$. In fact, the meniscus must be concave to make the driving Laplace pressure positive and the spontaneous imbibition possible. We will neglect the spherical cap and approximate a spherical meniscus by a flat one so that small corrections by the spherical cap to the liquid volume and the liquid-vapor surface area are neglected.
}
\label{fig:D2}
\end{figure}

We consider an axially symmetric capillary around the $z$ axis whose inlet is at $z=0$. We borrow the concept of transient state of the classical nucleation theory~\cite{Donne2022,Tinti2023}, and assume an imbibition pathway along the capillary with a constant Young's contact angle~\cite{Iwamatsu2020,Iwamatsu2022,Tinti2023}. Then, the solid-liquid surface free energy when the liquid-vapor interface reaches $z$ is given by~\cite{Iwamatsu2020}
\begin{eqnarray}
F_{\rm sl} &=& 2\pi \gamma_{\rm lv} \cos \theta_{\rm Y}  \int_{0}^{z} R(z')\sqrt{1+\left(\frac{dR}{dz'}\right)^{2}}dz',
\label{eq:D3}
\end{eqnarray}
where $R(z)$ is the radius of the capillary at $z$.  The liquid-vapor surface free energy is given by
\begin{equation}
F_{\rm lv} = 2\pi \gamma_{\rm lv}R(z)^2\frac{1-\cos\psi}{\sin^{2}\psi}\simeq  \pi \gamma_{\rm lv}R(z)^2,
\label{eq:D4}
\end{equation}
where the half-opening angle $\psi$ defined in Fig.~\ref{fig:D2} is approximated by $\psi=0$ to simplify mathematics. Therefore, a spherical interface is replaced by a flat one~\cite{Iwamatsu2020} because inclusion of the spherical interface gives only a small correction which can be included by regarding the capillary radius $R(z)$ as an effective radius~\cite{Iwamatsu2022}.

The total liquid volume inside the capillary is also approximately given by
\begin{equation}
V(z)=\pi\int_{0}^{z}R_{i}(z')^{2}dz'+\frac{\pi}{3}R(z)^{3}\nu\left(\psi\right)\simeq \pi\int_{0}^{z}R(z')^{2}dz'
\label{eq:D5}
\end{equation}
where a small volume correction of a spherical cap~\cite{Iwamatsu2022} (Fig.~\ref{fig:D2})
\begin{equation}
\nu\left(\psi\right)=\frac{\left(1-\cos\psi\right)^{2}\left(2+\cos\psi\right)}{\sin^{3}\psi}
\label{eq:D6}
\end{equation}
is neglected. Since the surface free energy $F(z)$ in Eq.~(\ref{eq:D1}) is simply given as a function of $z$ by the sum of Eqs.~(\ref{eq:D3}) and (\ref{eq:D4}), and the liquid volume $V(z)$ is given by Eq.~(\ref{eq:D5}), the liquid pressure $p_{\rm L}(z)$ defined by 
\begin{equation}
p_{\rm L}(z)=-\frac{\partial F}{\partial V}=-\frac{1}{dV/dz}\frac{\partial F_(z)}{\partial z},
\label{eq:D7}
\end{equation}
can be analytically calculated as
\begin{equation}
p_{\rm L}(z)=\frac{2\gamma_{\rm lv}}{R(z)}\left(\cos\theta_{\rm Y}\sqrt{1+\left(\frac{dR}{dz}\right)^{2}}-\frac{dR}{dz}\right)
\label{eq:D8}
\end{equation}
which reduces to the standard Laplace pressure
\begin{equation}
p_{\rm L}(z)=\frac{2\gamma_{\rm lv}\cos\theta_{\rm Y}}{R}
\label{eq:D9}
\end{equation}
in straight cylinders ($R(z)=R$, $dR/dz=0$).

\begin{figure}[htbp]
\begin{center}
\includegraphics[width=0.80\linewidth]{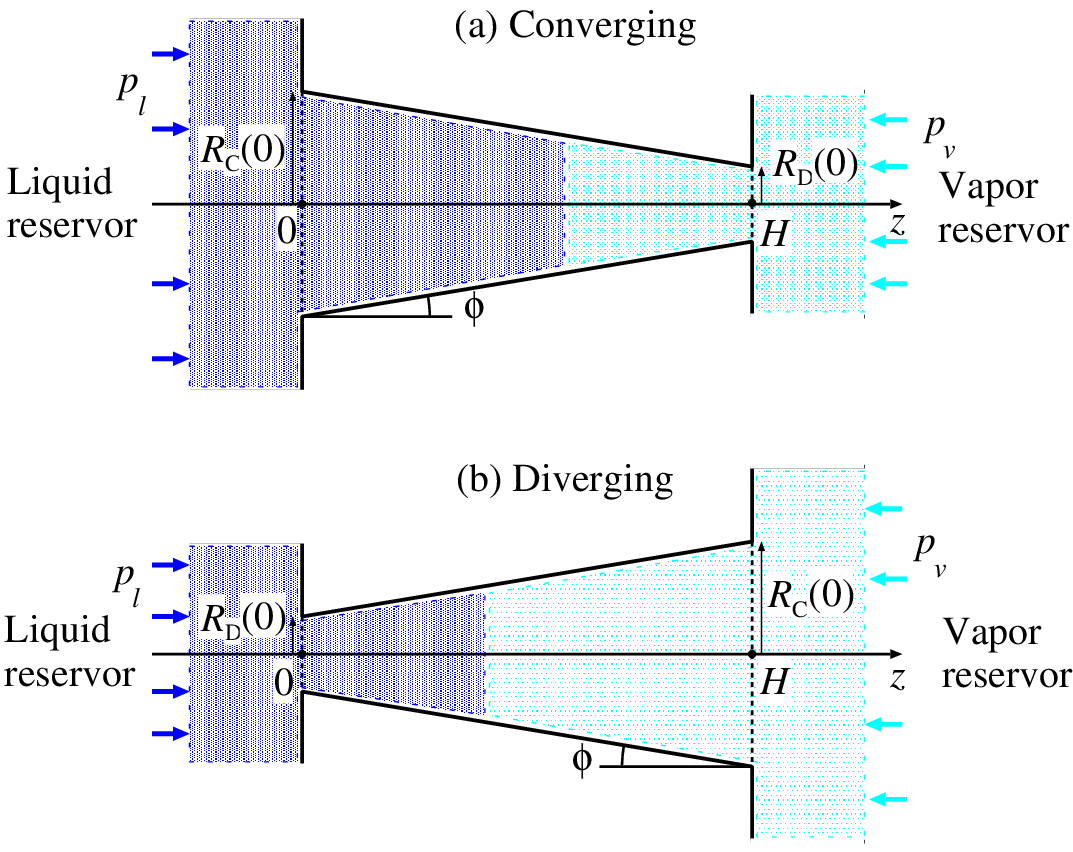}
\end{center}
\caption{
Two axially symmetric conical capillaries with (a) a converging radius and (b) a diverging radius. The inlet radius of the diverging capillary and that of the converging capillary are $R_{\rm C}(0)$ and $R_{\rm D}(0)$, respectively. The length of the capillary is $H$ and the tilt angle of wall is $\phi$. The inlet at the left is immersed in the liquid reservoir with the liquid pressure $p_{\rm l}$ and the outlet at the right is immersed in the vapor reservoir with the vapor pressure $p_{\rm v}$. The liquid intrudes from the left to the right. When the liquid intrusion occurs without applied external pressure $p_{\rm ext}=p_{\rm l}-p_{\rm v}=0$, the spontaneous imbibition (intrusion) is possible. The forced liquid intrusion and extrusion occurs by applying the external pressure $p_{\rm ext}\neq 0$. 
 }
\label{fig:D3}
\end{figure}

\begin{figure}[htbp]
\begin{center}
\includegraphics[width=0.8\linewidth]{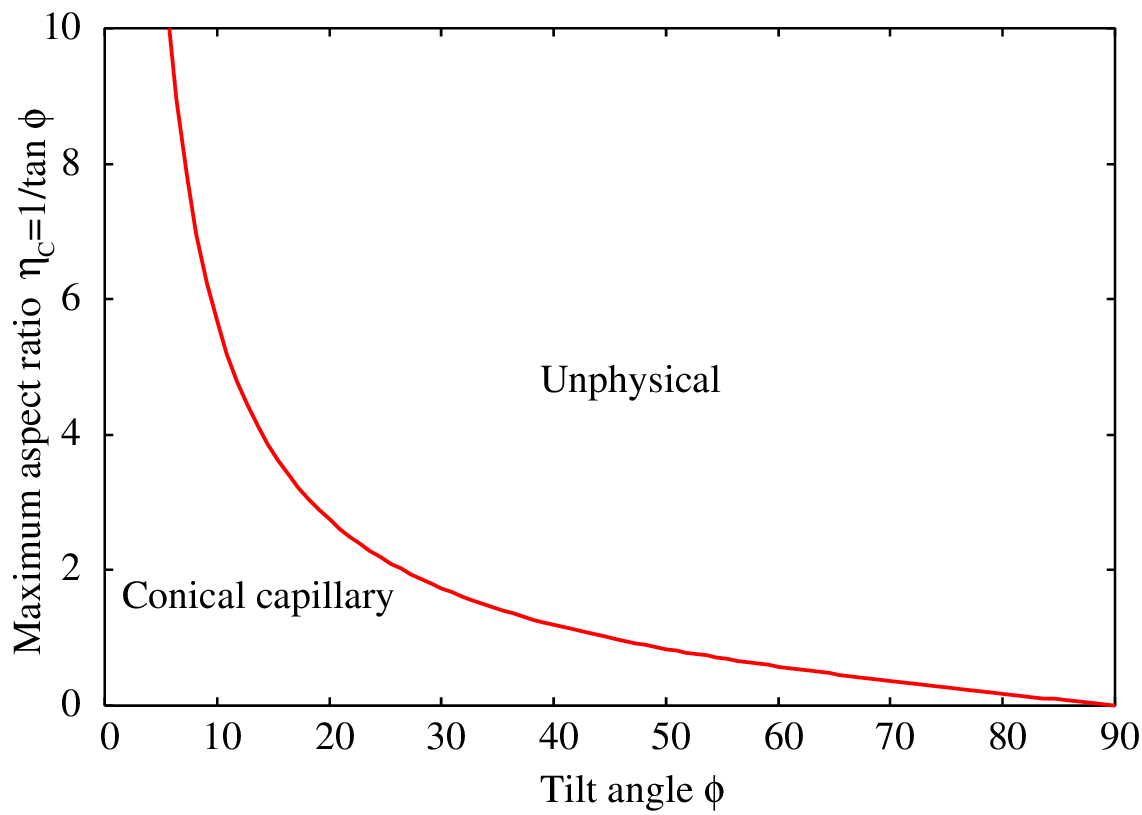}
\end{center}
\caption{
Maximum aspect ratio $\eta_{\rm C}=1/\tan\phi$ as a function of the tilt angle $\phi$.  Two fundamental parameters $\phi$ and $\eta_{\rm C}$ are not independent.  The region above the curve is not allowed.  A long capillary with high aspect ratio $\eta_{\rm C}=H/R_{\rm C}(0)$ is possible only when the tilt angle $\phi$ is low.
 } 
\label{fig:D4}
\end{figure}

Now, we consider two conical capillaries of identical shape with either a converging or a diverging radius (Fig.~\ref{fig:D3}) $R_{\rm C}(z)$ or $R_{\rm D}(z)$ given by:
\begin{eqnarray}
R_{\rm C}(z) = R_{\rm C}(0)-\left(\tan\phi\right) z,\;\;\;\left(0 \leq z\leq H\right),
\label{eq:D10} \\
R_{\rm D}(z) = R_{\rm D}(0)+\left(\tan\phi\right) z,\;\;\;\left(0 \leq z\leq H\right),
\label{eq:D11}
\end{eqnarray}
where $\phi (0\leq \phi\leq 90^{\circ})$, and $R_{\rm C}(0)=R_{\rm C}(z=0)$ and $R_{\rm D}(0)=R_{\rm D}(z=0)$, and $H$ represent the tilt angle of the wall, the radius at the inlet ($R_{\rm C}(0)> R_{\rm D}(0)$), and the length of the capillary (Fig.~\ref{fig:D3}).  Hence, the capillary parameters in Eqs.~(\ref{eq:D10}) and (\ref{eq:D11}) are related by
\begin{equation}
R_{\rm D}(0)=R_{\rm C}(0)-\left(\tan\phi\right) H.
\label{eq:D12}
\end{equation}
To study the imbibition, we have to specify the geometry of the conical capillaries. We select the tilt angle $\phi$ and the aspect ratio
\begin{equation}
\eta_{\rm C} = \frac{H}{R_{\rm C}(0)}
\label{eq:D13}
\end{equation}
as the two fundamental parameters to specify the geometry.  Therefore, another aspect ratio
\begin{equation}
\eta_{\rm D} =  \frac{H}{R_{\rm D}(0)}
\label{eq:D14}
\end{equation}
is determined from the two fundamental parameters by
\begin{equation}
\frac{1}{\eta_{\rm D}}=\frac{1}{\eta_{\rm C}}-\tan\phi
\label{eq:D15}
\end{equation}
from Eq.~(\ref{eq:D12}).  Furthermore, these two fundamental parameters are not independent owing to geometrical constraint $R_{\rm D}(0)\ge 0$ and they satisfy
\begin{equation}
0<\eta_{\rm C}\le \frac{1}{\tan\phi},
\label{eq:D16}
\end{equation}
where the equality holds when the capillary is a true cone with $R_{\rm D}(0)=0$. Figure \ref{fig:D4} presents the maximum aspect ratio $\eta_{\rm C}=1/\tan\phi$ for given tilt angle $\phi$.  A large aspect ratio $\eta_{\rm C}$ is possible only when the tilt angle $\phi$ is low.

The capillary pressure in Eq.~(\ref{eq:D8}) becomes a modified Laplace pressure written as
\begin{equation}
p_{{\rm L}(i)}(z)=\frac{2\gamma_{\rm lv}\Pi_{i}\left(\theta_{\rm Y},\phi\right)}{R_{i}(z)},
\label{eq:D17}
\end{equation}
where the index $i=$ C, D distinguishes the "Converging" and the "Diverging" geometry, and the scaled pressure $\Pi_{i}$ is given by
\begin{eqnarray}
\Pi_{\rm C}\left(\theta_{\rm Y},\phi\right) &=& \frac{\cos\theta_{\rm Y}}{\cos\phi}+\tan\phi,
\label{eq:D18} \\
\Pi_{\rm D}\left(\theta_{\rm Y},\phi\right) &=& \frac{\cos\theta_{\rm Y}}{\cos\phi}-\tan\phi
\label{eq:D19}
\end{eqnarray}
from Eqs.~(\ref{eq:D10}) and (\ref{eq:D11}), which determine the sign and the magnitude of the modified Laplace pressure in conical capillaries.

A more accurate pressure formula~\cite{Iwamatsu2022}, which takes into account the spherical liquid-vapor interface (Fig.~\ref{fig:D1}) was derived in our previous paper~\cite{Iwamatsu2022}, where the pore radius $R(z)$ in Eq.~(\ref{eq:D17}) is replace by an effective radius corrected by the small volume of spherical cap and the scaled pressures in Eqs.~(\ref{eq:D18}) and (\ref{eq:D19}) are replaced by~\cite{Iwamatsu2022}
\begin{eqnarray}
\Pi_{\rm C}\left(\theta_{\rm Y},\phi\right) &=& \frac{\cos\theta_{\rm Y}}{\cos\phi}+\frac{2\tan\phi}{1+\sin\left(\theta_{\rm Y}-\phi\right)},
\label{eq:D20} \\
\Pi_{\rm D}\left(\theta_{\rm Y},\phi\right) &=& \frac{\cos\theta_{\rm Y}}{\cos\phi}-\frac{2\tan\phi}{1+\sin\left(\theta_{\rm Y}+\phi\right)}.
\label{eq:D21} 
\end{eqnarray}
We can recover Eq.~(\ref{eq:D18}) and (\ref{eq:D19}) by setting $\psi=0$ (Fig.~\ref{fig:D1}) or $\theta_{\rm Y}-\phi=90^{\circ}$ in Eq.~(\ref{eq:D20}) and  $\theta_{\rm Y}+\phi=90^{\circ}$ in Eq.~(\ref{eq:D21}).

\begin{figure}[htbp]
\begin{center}
\includegraphics[width=0.8\linewidth]{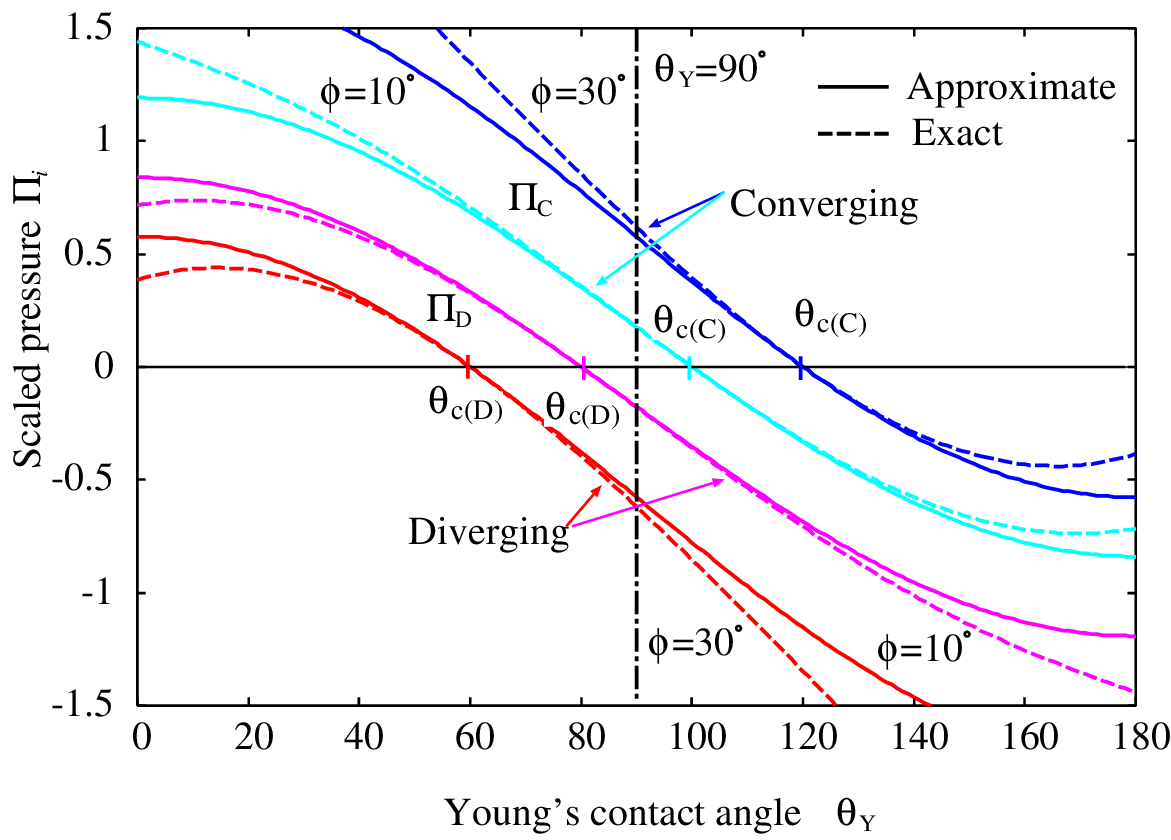}
\end{center}
\caption{
Scaled pressures $\Pi_{\rm C}\left(\theta_{\rm Y},\phi\right)$ and $\Pi_{\rm D}\left(\theta_{\rm Y},\phi\right)$ as functions of Young's contact angle $\theta_{\rm Y}$ for a low tilt angle $\phi=10^{\circ}$ and a high tilt angle $\phi=30^{\circ}$. The exact (Eqs.(\ref{eq:D20}) and (\ref{eq:D21}), broken lines) and the approximate (Eqs.~(\ref{eq:D18}) and (\ref{eq:D19}), solid lines) scaled pressure is compared. These two sets of curves are indistinguishable near the zeros at the critical Young's angles $\theta_{\rm c(C)}$ and $\theta_{\rm c(D)}$. Note that the critical angles of the converging capillary with $\phi=10^{\circ}$ and $30^{\circ}$ are $\theta_{\rm c(C)}=100^{\circ}$ and $120^{\circ}$, respectively, from Eq.~(\ref{eq:D22}). The critical angles of the diverging capillaries are $\theta_{\rm c(D)}=80^{\circ}$ and $60^{\circ}$, respectively, from Eq.~(\ref{eq:D23}). Spontaneous imbibition is possible when the scaled pressure is positive. Therefore, diode-like one-way transport only toward the converging direction but not toward the diverging direction~\cite{Iwamatsu2022} is realized when $\theta_{\rm c(D)}<\theta_{\rm Y}<\theta_{\rm c(C)}$.
} 
\label{fig:D5}
\end{figure}

Figure~\ref{fig:D5} presents the exact (Eqs.~(\ref{eq:D20}) and (\ref{eq:D21})) and the approximate (Eqs.~(\ref{eq:D18}) and (\ref{eq:D19})) scaled pressures $\Pi_{\rm C}\left(\theta_{\rm Y},\phi\right)$ and $\Pi_{\rm D}\left(\theta_{\rm Y},\phi\right)$ as a function of Young's contact angle $\theta_{\rm Y}$ for a low tilt angle $\phi=10^{\circ}$ and a high tilt angle $\phi=30^{\circ}$. Apparently, they have symmetry $\Pi_{\rm C}\left(\pi-\theta_{\rm Y},\phi\right)=-\Pi_{\rm D}\left(\theta_{\rm Y},\phi\right)$ and  $\Pi_{\rm C}\left(\theta_{\rm Y},-\phi\right)=\Pi_{\rm D}\left(\theta_{\rm Y},\phi\right)$.  An exact and an approximate curve does not differ appreciably unless the tilt angle $\phi$ and the Young's contact angle $\theta_{\rm Y}$ are high.  In particular, the two curves cross zero exactly at the same critical Young's angle $\theta_{\rm c(C)}$ and $\theta_{\rm c(D)}$, where the capillary pressure in Eq.~(\ref{eq:D17}) vanishes.  In order to keep our model as simple as possible, we will continue to use these approximate formulas in Eqs.~(\ref{eq:D4}) and (\ref{eq:D5}) which leads to Eqs.~(\ref{eq:D18}) and (\ref{eq:D19}).

Since the modified Laplace pressure $p_{{\rm L}(i)}(z)$ acts as the driving force of liquid intrusion, spontaneous intrusion is possible when the scaled pressure $\Pi_{i}\left(\theta_{\rm Y},\phi\right)$ is positive (see Fig.~\ref{fig:D5}).  This occurs when Young's angle $\theta_{\rm Y}$ is smaller than the critical Yong's angle $\theta_{{\rm c}(i)}$ ($\theta_{\rm Y}<\theta_{{\rm c}(i)}$).  Consequently, the imbibition in a converging capillary is possible but that in a diverging capillary is prohibited when $\theta_{\rm c(D)}<\theta_{\rm Y}<\theta_{\rm c(C)}$.  Therefore, a single conical capillary shows diode-like one-way transport: the intrusion into the converging direction will be realized but that into the diverging direction will be prohibited when $\theta_{\rm c(D)}<\theta_{\rm Y}<\theta_{\rm c(C)}$~\cite{Iwamatsu2022}.

These two critical Young's angles $\theta_{{\rm c}(i)}$ are determined from $\Pi_{i}\left(\theta_{{\rm c}(i)},\phi\right)=0$, which gives~\cite{Iwamatsu2020,Iwamatsu2022}
\begin{eqnarray}
\theta_{\rm c(C)} &=& 90^{\circ}+\phi,\;\;\;({\rm Converging}),
\label{eq:D22} \\
\theta_{\rm c(D)} &=& 90^{\circ}-\phi, \;\;\;({\rm Diverging}).
\label{eq:D23}
\end{eqnarray}
Geometrically, at this contact angle ($\theta_{\rm Y}=\theta_{{\rm c}(i)}$), the liquid-vapor meniscus becomes flat (see Fig.~\ref{fig:D2}) and the free-energy cost to move the meniscus vanishes because the free energy remain constant irrespective of the location of the meniscus position as $p_{{\rm L}(i)}=\partial F_{i}/\partial V_{i}=0$ at $\theta_{{\rm c}(i)}$. Then, the liquid-vapor interface will be delocalized and able to move freely. The liquid can fill or empty the capillary by the mechanism known as the filling transition of wedge and cone~\cite{Hauge1992,Reijmer1999,Malijevsky2015} even though the driving force of intrusion is absent ($p_{{\rm L}(i)}=0$).

\subsection{\label{sec:sec2.2}Spontaneous imbibition}

\begin{figure}[htbp]
\begin{center}
\includegraphics[width=0.8\linewidth]{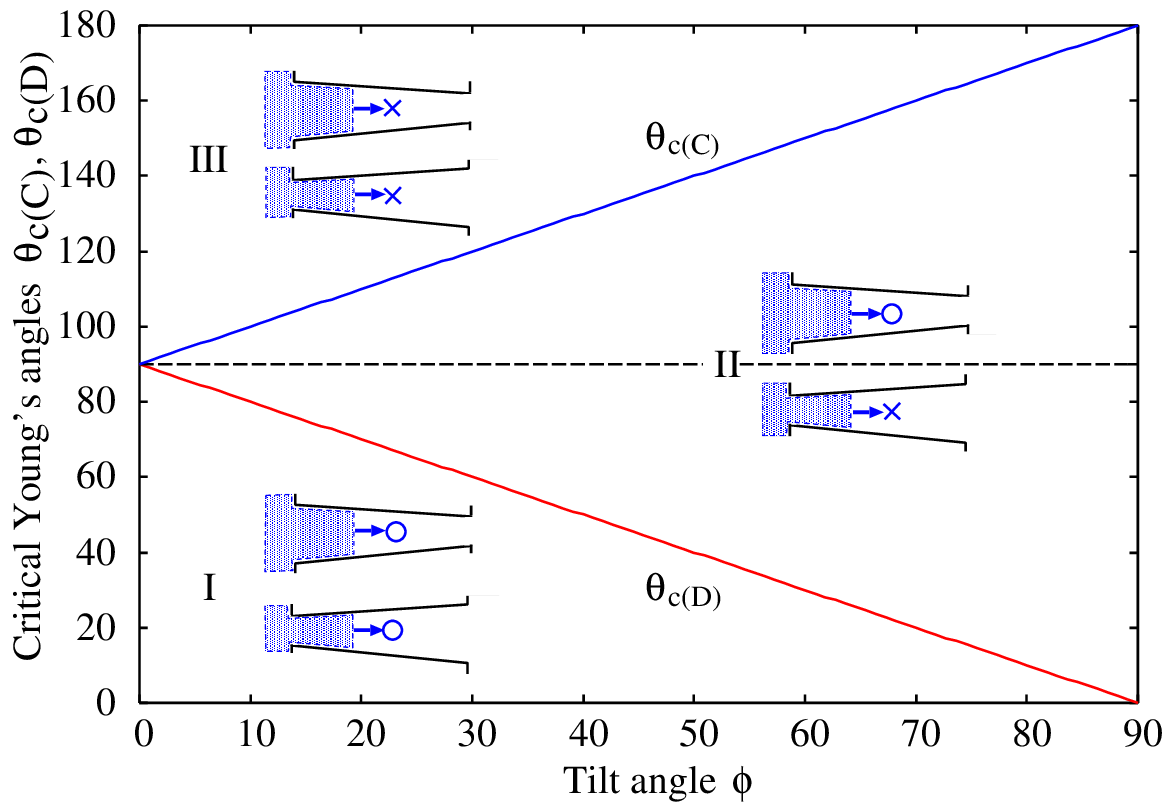}
\end{center}
\caption{
Critical Young's angles $\theta_{\rm c(C)}$ and $\theta_{\rm c(D)}$ given by Eqs.~(\ref{eq:D22}) and (\ref{eq:D23}) as functions of the tilt angles $\phi$. The $\left(\phi,\theta_{\rm Y}\right)$ space is divided into three regions I, II and III by these two lines.  In the region II, the spontaneous liquid intrusion is possible only in converging capillaries.  In the region I, the spontaneous liquid intrusion is possible in both converging capillaries and diverging capillaries, while it is possible in neither converging capillaries nor diverging capillaries in the region III.
 } 
\label{fig:D6}
\end{figure}

Figure~\ref{fig:D6} presents the critical Young's angles $\theta_{\rm c(C)}$ and $\theta_{\rm c(D)}$ as a function of the tilt angle $\phi$. The critical angle of converging capillary belongs to the "hydrophobic" region $\theta_{\rm c(C)}>90^{\circ}$, whereas that of the diverging one belongs to the "hydrophilic" region $\theta_{\rm c(D)}<90^{\circ}$. In this paper, we will use "hydrophobic" and "hydrophilic" instead of "lyophobic" and "lyophilic" though we will consider general liquid. Therefore, spontaneous imbibition of liquid can occur in converging capillaries even if they are hydrophobic ($90^{\circ}<\theta_{\rm Y}$) as long as $\theta_{\rm Y}<\theta_{\rm c(C)}$.

The two critical angles $\theta_{\rm c(C)}$ in Eq.~(\ref{eq:D22}) and $\theta_{\rm c(D)}$ in Eq~(\ref{eq:D23}) divide $\left(\phi,\theta_{\rm Y}\right)$ space in Fig.~\ref{fig:D6} into three regions I, II and III. In the region I, the spontaneous liquid intrusion occurs both in converging capillaries and in diverging capillaries because $\theta_{\rm Y}<\theta_{\rm c(D)}<\theta_{\rm c(C)}$. In the region II, the spontaneous intrusion only in converging capillaries is possible and that in diverging capillaries is prohibited because $\theta_{\rm c(D)}<\theta_{\rm Y}<\theta_{\rm c(C)}$. Therefore, in this region II, a single conical capillary functions as a liquid diode~\cite{Singh2020,Iwamatsu2022}. A larger tilt angle $\phi$ is advantageous from Fig.~\ref{fig:D6} to expand the region II. However, it requires a short capillary length $H$ with a low aspect ratio $\eta_{\rm C}$ from Fig.~\ref{fig:D4}.  In converging conical capillaries, furthermore, the spontaneous liquid intrusion can occur even if the wall is hydrophobic as long as $90^{\circ}<\theta_{\rm Y}< \theta_{\rm c(C)}$ or  $90^{\circ}<\theta_{\rm Y}< 90^{\circ}+\phi$ from Eq.~(\ref{eq:D22}). In fact, the spontaneous intrusion (infiltration) in hydrophobic and converging conical capillaries has been observed by the molecular dynamic simulation~\cite{Liu2009}.

In the region III, the spontaneous liquid intrusion is prohibited both in converging capillaries and in diverging capillaries because $\theta_{\rm c(D)}<\theta_{\rm c(C)}<\theta_{\rm Y}$.  Only the forced imbibition, which is realized by applying the external (infiltration) pressure (Fig.~\ref{fig:D3}) to liquid or vapor, is possible. Further investigation of the free energy landscape~\cite{Iwamatsu2020} is necessary to understand the details of the forced imbibition process.

\subsection{\label{sec:sec2.3}Free energy landscape of forced imbibition}

When the modified Laplace pressure in Eq.~(\ref{eq:D17}) is negative, i.e., $p_{{\rm L}(i)}(z)<0$ or $\Pi_{i}\left(\theta_{\rm Y},\phi\right)<0$, the spontaneous liquid intrusion is prohibited. It is necessary to apply a positive external pressure $p_{\rm ext}=p_{\rm l}-p_{\rm v}>0$ (Fig.~\ref{fig:D3}) to cancel this negative Laplace pressure to force the intrusion of liquid. On the other hand, when the capillary is completely filled by a positive Laplace pressure $p_{{\rm L}(i)}(z)>0$, it is necessary to apply a negative external pressure $p_{\rm ext}<0$ to cancel this positive Laplace pressure to force extrusion of liquid from the capillary. To determine the magnitude of the applied pressure $p_{\rm ext}$, we have to understand the free energy landscape of imbibition.

The thermodynamics of forced imbibition process is described by the free energy~\cite{Konig2004,Roth2006,Iwamatsu2020,Iwamatsu2022,Donne2022,Alzaidi2022}
\begin{equation}
\Omega_{i}=F_{i}-p_{\rm ext}V_{i},
\label{eq:D24}
\end{equation}
where $F_{i}$ is the surface free energy in Eq.~(\ref{eq:D1}) and $V_{i}$ is the liquid volume inside the capillary given by Eq.~(\ref{eq:D5}). Then, the driving capillary pressure becomes
\begin{equation}
p_{i}(z)=-\frac{\partial \Omega_{i}}{\partial V_{i}}=p_{\rm ext}+p_{{\rm L}(i)}(z),
\label{eq:D25}
\end{equation}
where $p_{{\rm L}(i)}(z)$ is the modified Laplace pressure in Eq.~(\ref{eq:D17}). If the driving pressure $p_{i}(z)$ is always positive within the capillary ($0\leq z \leq H$), the intrusion of liquid into the whole capillary is realized. If the driving pressure is always negative within the capillary, extrusion of liquid from the whole capillary is achieved, and the capillary will be empty and filled by vapor.

The liquid intrusion starts at the inlet ($z=0$) when $p_{i}(z=0)\ge 0$ in Eq.~(\ref{eq:D25}). The critical external pressure $p_{\rm ext}=p_{c(i)}$ is given by the condition $p_{i}(z)=0$ at $z=0$, which leads to 
\begin{equation}
p_{\rm{c}(i)}=-p_{{\rm L}(i)}(0)=-\frac{2\gamma_{\rm lv}\Pi_{i}\left(\theta_{\rm Y},\phi\right)}{R_{i}(0)}.
\label{eq:D26}
\end{equation}

To visualize the pressure and the free energy landscape, we introduce the non-dimensional pressure $\tilde{p}$ and the free energy $\omega_{i}$ through~\cite{Iwamatsu2020,Iwamatsu2022}
\begin{equation}\Omega_{i}=\gamma_{\rm lv}\pi R_{i}^{2}(0)\omega_{i}(\tilde{z}),
\label{eq:D27}
\end{equation}
and the non-dimensional quantities
\begin{equation}
\tilde{z} = \frac{z}{H},\;\;\;(0\leq \tilde{z}\leq 1)
\label{eq:D28}
\end{equation}
\begin{equation}
\tilde{p}=\frac{H p_{\rm ext}}{\gamma_{\rm lv}},
\;\;\;\alpha_{i}=\frac{H\tan\phi}{R_{i}(0)}=\eta_{i}\tan\phi.
\label{eq:D29}
\end{equation}
Therefore, the non-dimensional critical pressures are given by
\begin{eqnarray}
\tilde{p}_{\rm c(C)} &=& \frac{H p_{\rm c(C)}}{\gamma_{\rm lv}}=-2\eta_{\rm C}\Pi_{\rm C}\left(\theta_{\rm Y},\phi\right)
\label{eq:D30} \\
\tilde{p}_{\rm c(D)} &=& \frac{H p_{\rm c(D)}}{\gamma_{\rm lv}}=-2\eta_{\rm D}\Pi_{\rm D}\left(\theta_{\rm Y},\phi\right)
\label{eq:D31}
\end{eqnarray}
from Eq.~(\ref{eq:D26}).

\begin{figure}[htbp]
\begin{center}
\includegraphics[width=0.9\linewidth]{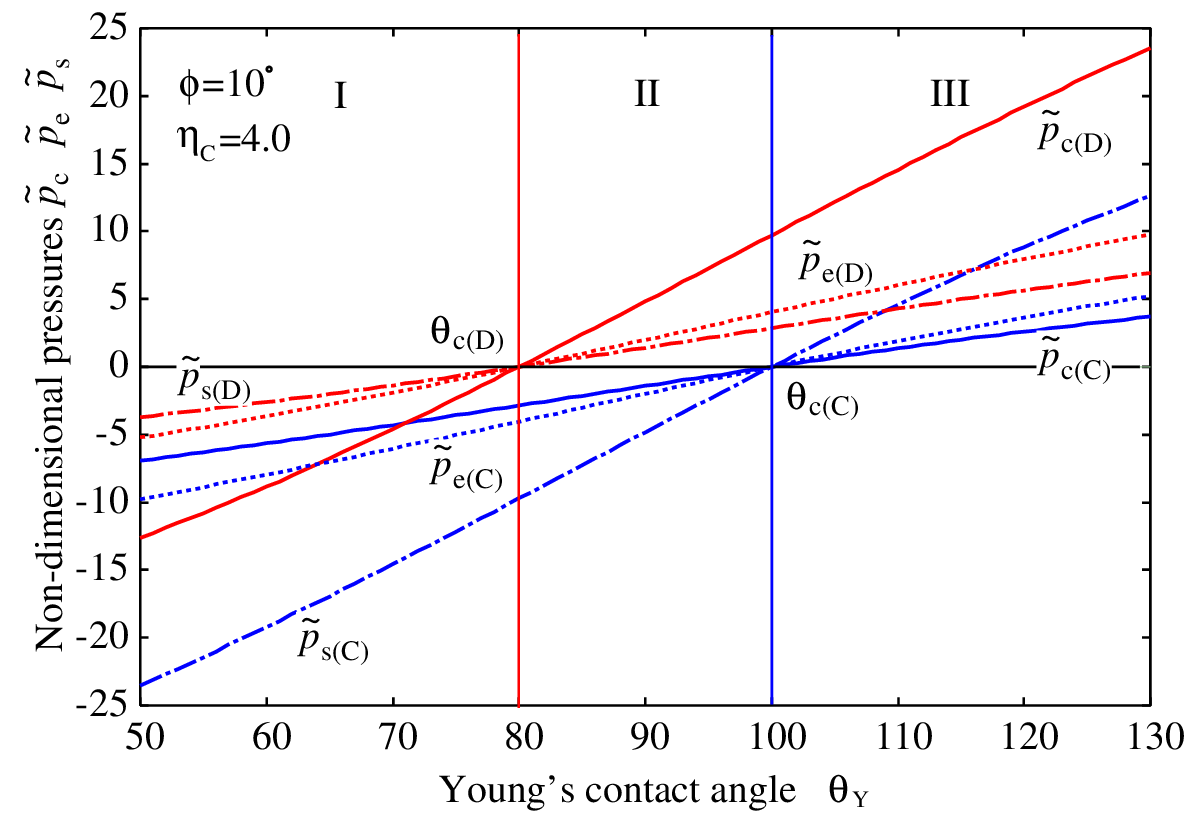}
\end{center}
\caption{
Non-dimensional characteristic pressures $\tilde{p}_{\rm c(C)}$, $\tilde{p}_{\rm e(C)}$, $\tilde{p}_{\rm s(C)}$ of a converging capillary and $\tilde{p}_{\rm c(D)}$, $\tilde{p}_{\rm e(D)}$, $\tilde{p}_{\rm s(D)}$ of a diverging capillary with $\phi=10^{\circ}$ and $\eta_{\rm C}=4.0$. Three regions III ($\theta_{\rm Y}>\theta_{\rm c(C)}$), II ($\theta_{\rm c(C)}>\theta_{\rm Y}>\theta_{\rm c(D)}$), and I ($\theta_{\rm Y}<\theta_{\rm c(D)}$) correspond to those in Fig.~\ref{fig:D6}. The critical angles are $\theta_{\rm c(C)}=90+10=100^{\circ}$ for the converging capillary and $\theta_{\rm c(D)}=90-10=80^{\circ}$ for the diverging capillary.  
 } 
\label{fig:D7}
\end{figure}

Figure~\ref{fig:D7} presents the non-dimensional critical pressure $\tilde{p}_{{\rm c}(i)}$ which corresponds to $p_{\rm{c}(i)}$ as a function of Young's angle $\theta_{\rm Y}$ when $\phi=10^{\circ}$ and $\eta_{\rm C}=4.0$. It also shows the two other characteristic pressures $\tilde{p}_{{\rm e}(i)}$ and $\tilde{p}_{{\rm s}(i)}$, whose meaning will be apparent soon. Note that this critical pressure $p_{c(i)}$ does not represent the entrance barrier pressure due to the potential barrier from atomic interactions~\cite{Mo2015}.

The free energy landscape $\Omega_{i}$ in Eq.~(\ref{eq:D24}) can be analytically calculated from Eqs.~(\ref{eq:D3})-(\ref{eq:D5}) and the non-dimensional free energy $\omega_{i}$ is given by cubic polynomials of $\tilde{z}$~\cite{Iwamatsu2020,Iwamatsu2022}:
\begin{eqnarray}
\omega_{\rm C}\left(\tilde{z}\right) &=& \left(\tilde{p}_{\rm c({\rm C})}-\tilde{p}\right)\tilde{z}
-\alpha_{\rm C}\left(\frac{\tilde{p}_{\rm c(C)}}{2}-\tilde{p}\right)\tilde{z}^{2}-\frac{1}{3}\alpha_{\rm C}^{2}\tilde{p}\tilde{z}^{3},
\nonumber \\
\label{eq:D32} \\
\omega_{\rm D}\left(\tilde{z}\right) &=& \left(\tilde{p}_{\rm c(D)}-\tilde{p}\right)\tilde{z}
+\alpha_{\rm D}\left(\frac{\tilde{p}_{\rm c(D)}}{2}-\tilde{p}\right)\tilde{z}^{2}-\frac{1}{3}\alpha_{\rm D}^{2}\tilde{p}\tilde{z}^{3}
\nonumber \\
\label{eq:D33}
\end{eqnarray}
for the converging and the diverging capillary, where we have dropped the constant free energy from the liquid vapor surface tension when the meniscus is located at the inlet ($\tilde{z}=0$).  Therefore, the origin of the free energy is always zero at $\tilde{z}=0$~\cite{Iwamatsu2020}.

\begin{figure}[htbp]
\begin{center}
\includegraphics[width=0.9\linewidth]{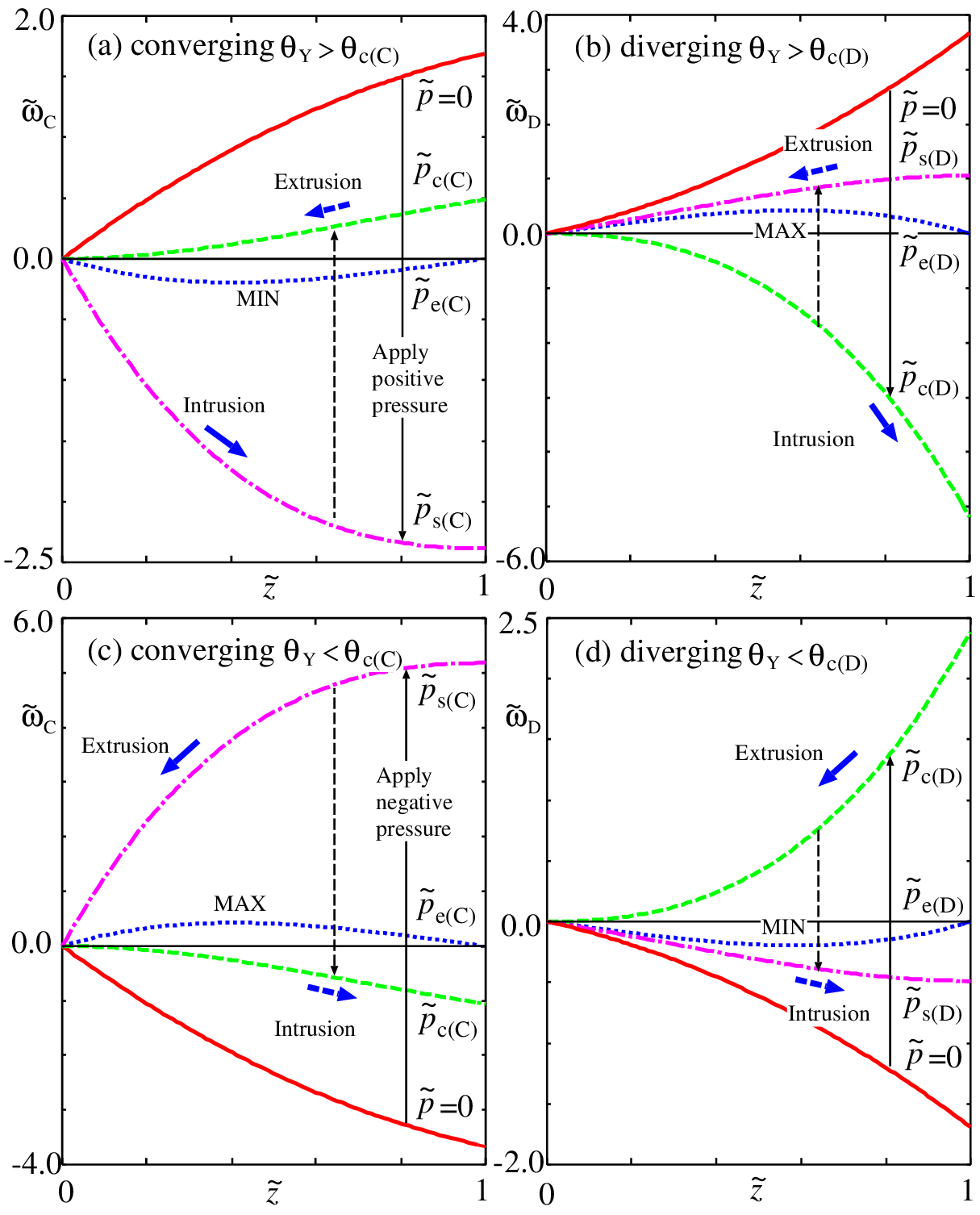}
\end{center}
\caption{
The free-energy landscape $\tilde{\omega}_{i}\left(\tilde{z}\right)$ of the liquid intrusion (infiltration) and the liquid extrusion in conical capillaries with $\phi=10^{\circ}$ and $\eta_{\rm C}=4.0$ at selected pressures $\tilde{p}=0$, $\tilde{p}_{{\rm c}(i)}$, $\tilde{p}_{{\rm e}(i)}$ and $\tilde{p}_{{\rm s}(i)}$ (Tab.~\ref{tab:D2}).  (a) The intrusion in a hydrophobic ($\theta_{\rm Y}=120^{\circ}$) and (c) the extrusion in a hydrophilic ($\theta_{\rm Y}=60^{\circ}$) converging capillary, and (b) the intrusion in a hydrophobic ($\theta_{\rm Y}=120^{\circ}$) and (d) the extrusion in a hydrophilic ($\theta_{\rm Y}=60^{\circ}$) diverging capillary. The liquid intrusion starts by increasing the magnitude of applied positive pressure (long thin solid down arrows) in (a) and (b), while the liquid extrusion occurs by increasing the magnitude of applied negative pressure (long thin solid up arrows) in (c) and (d). The vertical long thin broken arrows are reverse processes.  The free energy landscapes are characterized by either a maximum (MAX) which acts as a barrier or a minimum (MIN) which acts as a trap. They move from $\tilde{z}_{{\rm ex}(i)}=0$ at $\tilde{p}=\tilde{p}_{{\rm c}(i)}$ to $\tilde{z}_{{\rm ex}(i)}=1$ at $\tilde{p}=\tilde{p}_{{\rm s}(i)}$ from Eqs.~(\ref{eq:D37}) and (\ref{eq:D38}). The external pressure $\tilde{p}$ in the figure is fixed at $\tilde{p}=0$, $\tilde{p}_{{\rm c}(i)}$, $\tilde{p}_{{\rm e}(i)}$ and $\tilde{p}_{{\rm s}(i)}$ whose numerical values are tabulated in Tab.~\ref{tab:D2}.
} 
\label{fig:D8}
\end{figure}

Figure~\ref{fig:D8} presents the free energy landscape of imbibition along the pathway from $\tilde{z}=0$ (inlet) to $\tilde{z}=1$ (outlet), where the free energies are further scaled as
\begin{eqnarray}
\tilde{\omega}_{\rm C}\left(\tilde{z}\right) &=& \omega_{\rm C}\left(\tilde{z}\right)
\label{eq:D34} \\
\tilde{\omega}_{\rm D}\left(\tilde{z}\right) &=& \frac{R_{\rm D}^2(0)}{R_{\rm C}^2(0)}\omega_{\rm D}\left(\tilde{z}\right)=\left(1-\eta_{\rm C}\tan\phi\right)^2\omega_{\rm D}\left(\tilde{z}\right)
\label{eq:D35}
\end{eqnarray}
to make the scale of vertical axis (energy) common to both the converging capillary and the diverging capillary (see Eq.~(\ref{eq:D27})) since
\begin{equation}
\frac{R_{\rm D}^2(0)}{R_{\rm C}^2(0)}=\left(1-\eta_{\rm C}\tan\phi\right)^2
\label{eq:D36}
\end{equation}
from Eq.~(\ref{eq:D12}).

\begin{table}
\caption{
Non-dimensional characteristic pressures  $\tilde{p}_{{\rm c}(i)}$, $\tilde{p}_{{\rm e}(i)}$, $\tilde{p}_{{\rm s}(i)}$ for selected Young's angles $\theta_{\rm Y}$ used in Fig.~\ref{fig:D8} and Figs.~\ref{fig:D11}-\ref{fig:D14} for capillaries with $\phi=10^{\circ}$ and $\eta_{\rm C}=4.0$. 
}
\label{tab:D2}
\begin{tabular}{cc|cccccc}\hline
Region &  $\theta_{\rm Y}$ & $\tilde{p}_{\rm c(C)}$ & $\tilde{p}_{\rm e(C)}$ & $\tilde{p}_{\rm s(C)}$ &  $\tilde{p}_{\rm c(D)}$ & $\tilde{p}_{\rm e(D)}$ & $\tilde{p}_{\rm s(D)}$\\ \hline
 (I)  & $60^{\circ}$ & -5.66             & -7.96               &  -19.2              &  -8.85               & -3.67              & -2.61 \\ 
 (II) & $85^{\circ}$ & -2.14             & -3.01               &  -7.27              &  2.39               & 0.991              & 0.705 \\
 (II) & $95^{\circ}$ & -0.705             & -0.991               &  -2.39               &  7.27               & 3.01              & 2.14 \\ 
 (III) & $120^{\circ}$  & 2.61               & 3.67               &  8.85               &  19.2               & 7.96               & 5.66 \\
\hline
\end{tabular}
\end{table}

In Fig.~\ref{fig:D8}, we present the free energy landscapes at the selected pressures $\tilde{p}=0$ (spontaneous imbibition), $\tilde{p}_{{\rm c}(i)}$, $\tilde{p}_{{\rm e}(i)}$ and $\tilde{p}_{{\rm s}(i)}$.  The first characteristic pressure $\tilde{p}_{{\rm c}(i)}$ given by Eqs.~(\ref{eq:D30}) and (\ref{eq:D31}) is the critical pressure which characterizes the onset of imbibition at the inlet ($d\tilde{\omega}/d\tilde{z}|_{\tilde{z}=0}=0$ at $z=0$). The second characteristic pressure $\tilde{p}_{{\rm e}(i)}$ are given by~\cite{Iwamatsu2020,Iwamatsu2022}
\begin{eqnarray}
\tilde{p}_{\rm e(C)} &=& \frac{3\left(2-\alpha_{\rm C}\right)}{2\left(3-3\alpha_{\rm C}+\alpha_{\rm C}^{2}\right)}\tilde{p}_{\rm c(C)}.
\label{eq:D43} \\
\tilde{p}_{\rm e(D)} &=& \frac{3\left(2+\alpha_{\rm D}\right)}{2\left(3+3\alpha_{\rm D}+\alpha_{\rm D}^{2}\right)}\tilde{p}_{\rm c(D)}.
\label{eq:D44} 
\end{eqnarray}
where the free energy of the completely empty state and that of the completely filled state becomes equal $\tilde{\omega}\left(\tilde{z}=0\right)=\tilde{\omega}\left(\tilde{z}=1\right)=0$ (see Fig.~\ref{fig:D8}). This condition is similar to the two-phase coexistence of first-order phase transition.

The third characteristic pressure $\tilde{p}_{{\rm s}(i)}$ given by~\cite{Iwamatsu2020,Iwamatsu2022}
\begin{eqnarray}
\tilde{p}_{\rm s(C)} &=& \frac{\tilde{p}_{\rm c(C)}}{1-\alpha_{\rm C}},
\label{eq:D37} \\
\tilde{p}_{\rm s(D)} &=& \frac{\tilde{p}_{\rm c(D)}}{1+\alpha_{\rm D}},
\label{eq:D38}
\end{eqnarray}
characterize the stability limit of intruded liquid at the outlet ($d\tilde{\omega}/d\tilde{z}|_{\tilde{z}=1}=0$ at $z=H$) when the liquid starts to flow out from the outlet.  In fact, they correspond simply to the modified Laplace pressure at the outlet ($z=H$):
\begin{eqnarray}
p_{\rm s(C)} &=& -p_{\rm L(C)}(H)=-\frac{2\gamma_{\rm lv}\Pi_{\rm C}\left(\theta_{\rm Y},\phi\right)}{R_{\rm C}(H)},
\label{eq:D39} \\
p_{\rm s(D)} &=& -p_{\rm L(D)}(H)=-\frac{2\gamma_{\rm lv}\Pi_{\rm D}\left(\theta_{\rm Y},\phi\right)}{R_{\rm D}(H)},
\label{eq:D40}
\end{eqnarray}
in the original unit from Eqs,~(\ref{eq:D10}), (\ref{eq:D11}) and (\ref{eq:D17}). Therefore, the characteristic pressures $\tilde{p}_{{\rm c}(i)}$ and $\tilde{p}_{{\rm s}(i)}$ of the converging and the diverging capillary are related by
\begin{eqnarray}
p_{\rm s(C)} &=& \frac{\Pi_{\rm C}\left(\theta_{\rm Y},\phi\right)}{\Pi_{\rm D}\left(\theta_{\rm Y},\phi\right)}p_{\rm c(D)}.
\label{eq:D41} \\
p_{\rm s(D)} &=& \frac{\Pi_{\rm D}\left(\theta_{\rm Y},\phi\right)}{\Pi_{\rm C}\left(\theta_{\rm Y},\phi\right)}p_{\rm c(C)}.
\label{eq:D42}
\end{eqnarray}
because $R_{\rm C}(H)=R_{\rm D}(0)$ and $R_{\rm D}(H)=R_{\rm C}(0)$ so that $\tilde{p}_{\rm s(C)}$ and $\tilde{p}_{\rm c(D)}$, and $\tilde{p}_{\rm s(D)}$ and $\tilde{p}_{\rm c(C)}$ run almost in parallel in Fig.~\ref{fig:D7}.

When the external pressure $\tilde{p}$ is between $\tilde{p}_{{\rm c}(i)}$ and $\tilde{p}_{{\rm s}(i)}$ (e.g. Fig.~\ref{fig:D8} for $\tilde{p}=\tilde{p}_{{\rm e}(i)}$), the free energy landscape exhibits an extremum at $\tilde{z}_{{\rm ex}(i)}$ given by~\cite{Iwamatsu2020}
\begin{eqnarray}
\tilde{z}_{\rm ex(C)} &=& \frac{1}{\alpha_{\rm C}}\left(1-\frac{\tilde{p}_{\rm c(C)}}{\tilde{p}}\right),
\label{eq:D45} \\
\tilde{z}_{\rm ex(D)} &=& -\frac{1}{\alpha_{\rm D}}\left(1-\frac{\tilde{p}_{\rm c(D)}}{\tilde{p}}\right),
\label{eq:D46}
\end{eqnarray}
which move from $\tilde{z}_{{\rm ex}(i)}=0$ at $\tilde{p}=\tilde{p}_{{\rm c}(i)}$ to $\tilde{z}_{{\rm ex}(i)}=1$ at $\tilde{p}=\tilde{p}_{{\rm s}(i)}$ from Eqs.~(\ref{eq:D37}) and (\ref{eq:D38}), and their free energies become~\cite{Iwamatsu2020}
\begin{eqnarray}
\omega_{\rm ex(C)} &= &-\frac{\left(\tilde{p}-\tilde{p}_{\rm c(C)}\right)^{2}\left(2\tilde{p}+\tilde{p}_{\rm c(C)}\right)}{6\alpha_{\rm C}\tilde{p}^{2}},
\label{eq:D47} \\
\omega_{\rm ex(D)} &= &\frac{\left(\tilde{p}-\tilde{p}_{\rm c(D)}\right)^{2}\left(2\tilde{p}+\tilde{p}_{\rm c(D)}\right)}{6\alpha_{\rm D}\tilde{p}^{2}},
\label{eq:D48}
\end{eqnarray}
from Eqs.~(\ref{eq:D32}) and (\ref{eq:D33}), which correspond to a minimum when $\omega_{{\rm ex}(i)}<0$ (Figs.~\ref{fig:D8}(a) and (d)) and a maximum when $\omega_{{\rm ex}(i)}>0$  (Figs.~\ref{fig:D8}(b) and (c)).   From Eqs.~(\ref{eq:D27}), (\ref{eq:D47}) and (\ref{eq:D48}), and  noting $\tilde{p}\sim O(1)$ from Tab.~\ref{tab:D2}, we have roughly
\begin{equation}
\left| \Omega_{{\rm ex}(i)} \right| =\gamma_{\rm lv}\pi R_{i}^{2}(0)\left| \omega_{{\rm ex}(i)}\right| \simeq
\frac{\gamma_{\rm lv}\pi R_{i}^{2}(0)}{\alpha_{i}}=\frac{\gamma_{\rm lv}\pi R_{i}^{3}(0)}{H\tan\phi},
\label{eq:D49}
\end{equation}
which gives, for example, $\Omega_{{\rm ex}(i)}\simeq 1.3\times 10^{-15}$ J  $\gg kT\sim  4.0\times 10^{-21}$ J  for water ($\gamma_{\rm lv}=0.072$ J/m$^{2}$) in $R_{i}(0)=0.1$ $\mu$m, $H=1.0$ $\mu$m, and $\phi=10^{\circ}$ capillary. Of course, the thermal fluctuation $kT$ may not be negligible when the size of capillary is reduced to 1/100 ($R(0)=1$ nm and $H=10$ nm).

In Fig.~\ref{fig:D8}, the free energy landscapes of a converging capillary with $\phi=10^{\circ}$ and $\eta_{\rm C}=4.0$ are presented in Figs.~\ref{fig:D8}(a) and (c), and those of a diverging capillary are presented in Figs.~\ref{fig:D8}(b) and (d).  The numerical values of $\tilde{p}_{{\rm c}(i)}$, $\tilde{p}_{{\rm e}(i)}$ and $\tilde{p}_{{\rm s}(i)}$ used are tabulated in Tab.~\ref{tab:D2}, and $\alpha_{\rm C}\simeq 0.705$ and $\alpha_{\rm D}\simeq 2.393$ from Eq.~(\ref{eq:D29}).  In original unit, $p_{\rm ext}=\gamma_{\rm lv}\tilde{p}/H$ from Eq.~(\ref{eq:D29}). So, for example, $\tilde{p}=1.0$ corresponds to $p_{\rm ext}=7.2\times 10^{4}$ Pa for water in $H=1.0$ $\mu$m capillary.

The wettability of capillary in Figs.~\ref{fig:D8}(a) and (b) is hydrophobic with $\theta_{\rm Y}=120^{\circ}$ so that the spontaneous liquid intrusion is prohibited. Hence, the landscapes presented in Figs.~\ref{fig:D8}(a) and (b) represent the pathway of {\it liquid intrusion} by positive applied pressures which are termed intrusion or infiltration pressure~\cite{Liu2009,Goldsmith2009,Mo2015}. The wettability of capillary in Figs.~\ref{fig:D8}(c) and (d) is hydrophilic with $\theta_{\rm Y}=60^{\circ}$ so that the capillary is filled by spontaneously intruded liquid. Hence, the landscapes presented in Figs.~\ref{fig:D8}(c) and (d) represent the pathway of the {\it liquid extrusion} or the {\it vapor intrusion} when the applied pressure is {\it negative}~\cite{Iwamatsu2020}. That is when the liquid pressure is lower than the vapor pressure. The free energy landscape of the initial state at $\tilde{p}=0$ is that of the spontaneous imbibition and is monotonically increasing (Figs.~\ref{fig:D8}(a) and (b)) or decreasing (Figs.~\ref{fig:D8}(c) and (d)) function indicating the complete liquid extrusion or intrusion.

\begin{figure}[htbp]
\begin{center}
\includegraphics[width=0.9\linewidth]{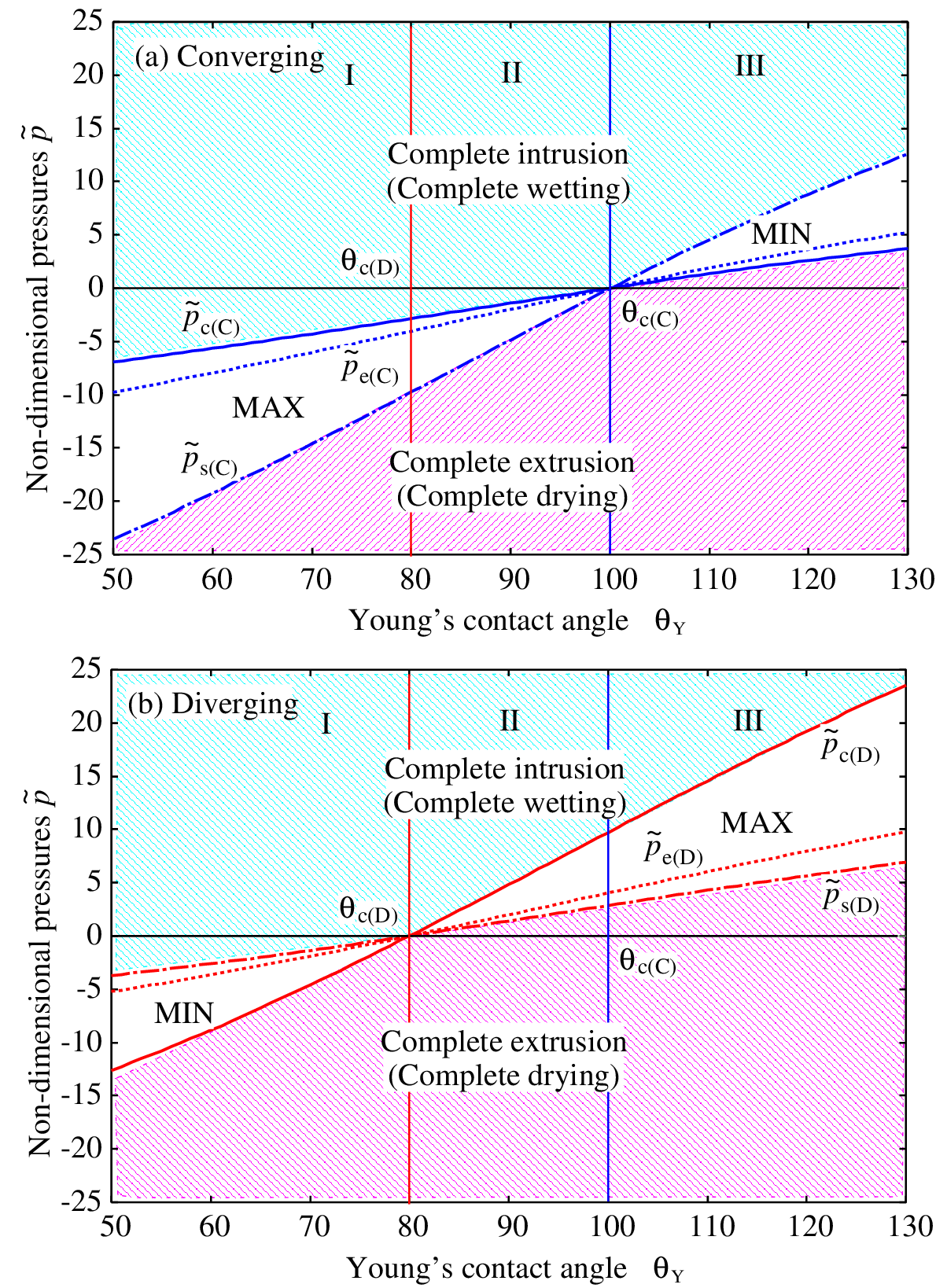}
\end{center}
\caption{
Wetting phase diagrams based on Fig.~\ref{fig:D7} in (a) converging and (b) diverging conical capillaries with $\phi=10^{\circ}$ and $\eta_{\rm C}=4.0$. These geometrical parameters are the same as those used in Figs.~\ref{fig:D7} and \ref{fig:D8}. At high pressures, the liquid completely intrudes into the capillary (complete wetting), while at low pressures the liquid completely extrudes from the capillary (complete drying). The free energy maximum (MAX) which acts as the free energy barrier or the minimum (MIN) which acts as the free energy minimum appears between $\tilde{p}_{{\rm s}(i)}$ and $\tilde{p}_{{\rm c}(i)}$ (Fig.~\ref{fig:D8}). The MAX suggests the pressure-induced first-order like wetting transition while the MIN suggests the second-order like continuous transition
} 
\label{fig:D9}
\end{figure}

This free energy landscape of the intrusion and the extrusion in Fig.~\ref{fig:D8} can be interpreted as that of the wetting and the drying transition~\cite{Dietrich1988,Rauscher2008} of complete wetting and drying in conical capillaries induced by the external applied pressure.  Figure~\ref{fig:D9} presents the phase diagrams of a converging and a diverging conical capillary. The complete intrusion corresponds to the complete wetting and the complete extrusion corresponds to the complete drying.  Therefore, the characteristic pressure $\tilde{p}_{{\rm e}(i)}$ corresponds to the "binodal" and the two critical pressures $\tilde{p}_{{\rm c}(i)}$ and $\tilde{p}_{{\rm s}(i)}$ correspond to the upper (lower) and the lower (upper) "spinodals" in the language of wetting transition.  The free energy landscape with a maximum, which acts as a free energy barrier, indicates the first order-like wetting and drying transition with pressure hysteresis and meniscus jumps, while that with a minimum indicates the second order-like transition with continuous change of the meniscus position.  In the former case, the meniscus trapped at the inlet (complete drying state) or at the outlet (complete wetting state) separated by the free energy barrier can be in the "metastable" state so that it can be destroyed and transition (intrusion or extrusion) can be induced by any perturbation.

It is relatively easy to imagine the scenario of intrusion and extrusion from the phase diagram in Fig.~\ref{fig:D9}.  However, to understand more complex scenarios in double-conical capillaries in the next section, it is helpful to look into the details of the intrusion and the extrusion processes from the free energy landscapes in Fig.~\ref{fig:D8}(a)-(d), which are summarize in Tab.~\ref{tab:D3}(a)-(d) and are as follows:

\renewcommand{\labelenumi}{(\theenumi)}
\begin{enumerate}
\item 
Figure~\ref{fig:D8}(a) presents the free energy landscape of the {\it liquid intrusion} or {\it infiltration}~\cite{Liu2009,Mo2015} in a hydrophobic ($\theta_{\rm Y}>\theta_{\rm c(C)}$) converging capillary. By increasing the (positive) applied pressure from $\tilde{p}=0$ (a long thin solid down arrow in Fig.~\ref{fig:D8}(a)), the intrusion starts at $\tilde{p}_{\rm c(C)}$ (Eq.~(\ref{eq:D30})) and is completed at $\tilde{p}_{\rm s(C)}$ (Eq.~(\ref{eq:D37})).  When $\tilde{p}_{\rm c(C)}\leq \tilde{p}\leq \tilde{p}_{\rm s(C)}$ (Fig.~\ref{fig:D9}(a)), the landscape shows a {\it minimum} (MIN), which acts as the {\it trap} for the liquid-vapor meniscus.  The location of MIN moves continuously from the inlet ($\tilde{z}=0$) toward the outlet ($\tilde{z}=1$).  Therefore, the intrusion occurs {\it gradually} (a short thick solid right arrow in Fig.~\ref{fig:D8}(a)).  In fact, such a gradual motion of meniscus is observed in the molecular dynamic simulation~\cite{Liu2009}.  In addition, the reverse process of liquid extrusion  (a long thin broken up arrow and a short thick broken left arrow in Fig.~\ref{fig:D8}(a)) is also gradual and completed at $\tilde{p}_{\rm c(C)}$ (see Tab.~\ref{tab:D3}(a)).

\item 
Figure~\ref{fig:D8}(b) presents the landscape of the {\it liquid intrusion} in a hydrophobic ($\theta_{\rm Y}>\theta_{\rm c(D)}$) diverging capillary  (Fig.~\ref{fig:D9}(b)).  The free energy landscape shows a {\it maximum} (MAX), which acts as the {\it barrier}. As the pressure is increased from $\tilde{p}_{\rm s(D)}$ towards $\tilde{p}_{\rm c(D)}$ (a long thin solid down arrow in Fig.~\ref{fig:D8}(b)), the meniscus is trapped at the inlet and is in metastable state, while the location of MAX moves from the outlet ($\tilde{z}=1$) toward the inlet ($\tilde{z}=0$). Therefore, the intrusion occurs {\it abruptly} at $\tilde{p}_{\rm c(D)}$ when the barrier reaches the inlet and disappears. The meniscus jumps from the inlet to the outlet at $\tilde{p}_{\rm c(D)}$ (a short thick solid right arrow in Fig.~\ref{fig:D8}(b)). Hence, this hydrophobic diverging capillary could act as a pneumatic switch~\cite{Preston2019}. Of course, any perturbations, such as mechanical vibration, thermal fluctuation may assist the meniscus to overcome the barrier estimated in Eq.~(\ref{eq:D49}). In the reverse process of depressing $\tilde{p}$ from $\tilde{p}_{\rm c(D)}$ to $\tilde{p}_{\rm s(D)}$ (a long thin broken up arrow in Fig.~\ref{fig:D8}(b)), the meniscus jumps from the outlet to the inlet at $\tilde{p}_{\rm s(D)}$ (a short thick broken left arrow in Fig.~\ref{fig:D8}(b)). Therefore, we will observe a pressure hysteresis between $\tilde{p}_{\rm c(D)}$ and $\tilde{p}_{\rm s(D)}$ (Tab.~\ref{tab:D3}(b)).

\item 
Figure~\ref{fig:D8}(c) presents the landscape of the {\it liquid extrusion} or the {\it vapor intrusion} in a hydrophilic ($\theta_{\rm Y}<\theta_{\rm c(C)}$) converging capillary. By increasing the absolute magnitude of negative applied pressure from $\tilde{p}=0$ (a long thin solid up arrow in Fig.~\ref{fig:D8}(c)), the complete extrusion occurs at $\tilde{p}_{\rm s(C)}$ (<0, Tab.~\ref{tab:D2}, Fig.~\ref{fig:D9}(a)). The free energy landscape shows a {\it maximum} (MAX). Therefore, the extrusion occurs {\it abruptly}. Since the extrusion of liquid from a converging capillary is the intrusion of vapor into a diverging capillary, this process in Fig.~\ref{fig:D8}(c) is similar to that in Fig.~\ref{fig:D8}(b). The reverse process of the complete liquid intrusion by depression (a long thin broken down arrow in Fig.~\ref{fig:D8}(c)) occurs also {\it abruptly} (a short thick broken right arrow in Fig.~\ref{fig:D8}(c)) at $\tilde{p}_{\rm c(C)}$ (<0). Again, we will observe a pressure hysteresis between $\tilde{p}_{\rm c(C)}$ and $\tilde{p}_{\rm s(C)}$ (Tab.~\ref{tab:D3}(c)) .

\item 
Figure~\ref{fig:D8}(d) presents the landscape of the {\it liquid extrusion} in a hydrophilic ($\theta_{\rm Y}<\theta_{\rm c(D)}$) diverging capillary. The extrusion occurs {\it gradually} as the landscape shows a {\it minimum} (MIN), and is completed at $\tilde{p}_{\rm c(D)}$. This process in Fig.~\ref{fig:D8}(d) is similar to that in Fig.~\ref{fig:D8}(a). The reverse process of the complete liquid intrusion is also gradual and is completed at $\tilde{p}_{\rm s(D)}$ (Tab.~\ref{tab:D3}(d)).

\end{enumerate}

\begin{table}
\caption{
Imbibition processes in a converging and a diverging conical capillary in Fig.~\ref{fig:D8}, which connects an empty state [E] and a filled state [F], where the down arrow indicates the compression and the up arrow indicates the depression.  Numerical values of the characteristic pressures $\tilde{p}_{\rm c(C)}$, $\tilde{p}_{\rm c(D)}$, $\tilde{p}_{\rm s(C)}$ and $\tilde{p}_{\rm s(D)}$ are tabulated in Tab.~\ref{tab:D2}.  Two letters "a" and "g" beside the arrows indicate that imbibition occurs {\it abruptly} (a) or {\it gradually} (g). Abrupt changes always accompany pressure hysteresis.
}
\label{tab:D3}
\begin{tabular}{cc|cc}\hline\hline
(a)                                    &                                            & (b)                                 &  \\
Intrusion                          & Reverse                                 & Intrusion                        & Reverse \\
\hline
$\tilde{p}=0$ [E]              & $\tilde{p}_{\rm c(C)}$ [E]          & $\tilde{p}=0$ [E]             & $\tilde{p}_{\rm s(D)}$ [E] \\
$\downarrow$g               & $\uparrow$g                           & $\downarrow$a              & $\uparrow$a \\
$\tilde{p}_{\rm s(C)}$ [F]  & $\tilde{p}_{\rm s(C)}$ [F]          & $\tilde{p}_{\rm c(D)}$ [F]  & $\tilde{p}_{\rm c(D)}$ [F]\\
\hline\hline
(c)                                &                                               & (d)                                 &  \\
Extrusion                       & Reverse                                  & Extrusion                       & Reverse \\
\hline
$\tilde{p}_{\rm s(C)}$ [E] & $\tilde{p}_{\rm s(C)}$ [E]          & $\tilde{p}_{\rm c(D)}$ [E]   & $\tilde{p}_{\rm c(D)}$ [E] \\
$\uparrow$a                  & $\downarrow$a                       & $\uparrow$g                    & $\downarrow$g \\
$\tilde{p}=0$ [F]            &  $\tilde{p}_{\rm c(C)}$ [F]         & $\tilde{p}=0$ [F]               & $\tilde{p}_{\rm s(D)}$ [F] \\
\hline\hline
\end{tabular}
\end{table}

The external pressure necessary to complete the intrusion and the extrusion can be predicted simply from the highest Laplace pressure, which is achieved either at the inlet or the outlet where the cross section is narrowest in conical capillaries. However, the intrusion and the extrusion process can be either {\it gradual} or {\it abrupt} depending on the shape of the free energy landscape.  They occur {\it gradually} if the free energy landscape has a {\it minimum} (Figs.~\ref{fig:D8}(a) and (d)), while they occur {\it abruptly} if the landscape has a {\it maximum} (Figs.~\ref{fig:D8}(b) and (c)).  We note again that this free energy maximum is not the nucleation barrier of critical droplet or bubble nucleated in the middle of capillaries~\cite{Lefevre2004,Remsing2015,Tinti2017,Tinti2023}.

The free energy maximum, which occurs in a hydrophobic diverging capillary (Fig.~\ref{fig:D8}(b)) and a hydrophilic converging capillary (Fig.~\ref{fig:D8}(c)), always accompanies a pressure hysteresis:
\begin{eqnarray}
\Delta p_{\rm hyst} &=& p_{\rm c(C)}-p_{\rm s(C)},\;\;\;({\rm converging}),
\label{eq:D50} \\
                            &=& p_{\rm c(D)}-p_{\rm s(D)},\;\;\;({\rm diverging}),
\label{eq:D51}
\end{eqnarray}
or
\begin{equation}
\Delta p_{\rm hyst}=2\gamma_{\rm lv}\left|\Pi_{\rm i}\left(\theta_{\rm Y},\phi\right)\right|\left(\frac{1}{R_{\rm C}(0)}-\frac{1}{R_{\rm D}(0)}\right)
\label{eq:D52}
\end{equation}
from Eqs.~(\ref{eq:D26}), (\ref{eq:D39}) and (\ref{eq:D40}).  Therefore, the pressure hysteresis is simply the difference between the modified Laplace pressure at the inlet and that at the outlet, or between the highest and the lowest modified Laplace pressure.  Based on these scenarios in Fig.~\ref{fig:D8} and Tab.~\ref{tab:D3}, we will discuss the intrusion and the extrusion in double conical capillaries in the next section.

\section{\label{sec:sec3}Imbibition in double conical capillaries}

\subsection{\label{sec:sec3.1}Spontaneous imbibition}

Based on the knowledge of the imbibition in a single conical capillary, we consider the intrusion and the extrusion in double conical capillaries. Specifically, we consider four capillaries: hourglass, diamond, sawtooth-1, and sawtooth-2 shaped capillaries illustrated in Fig.~\ref{fig:D1}.

\begin{figure}[htbp]
\begin{center}
\includegraphics[width=0.9\linewidth]{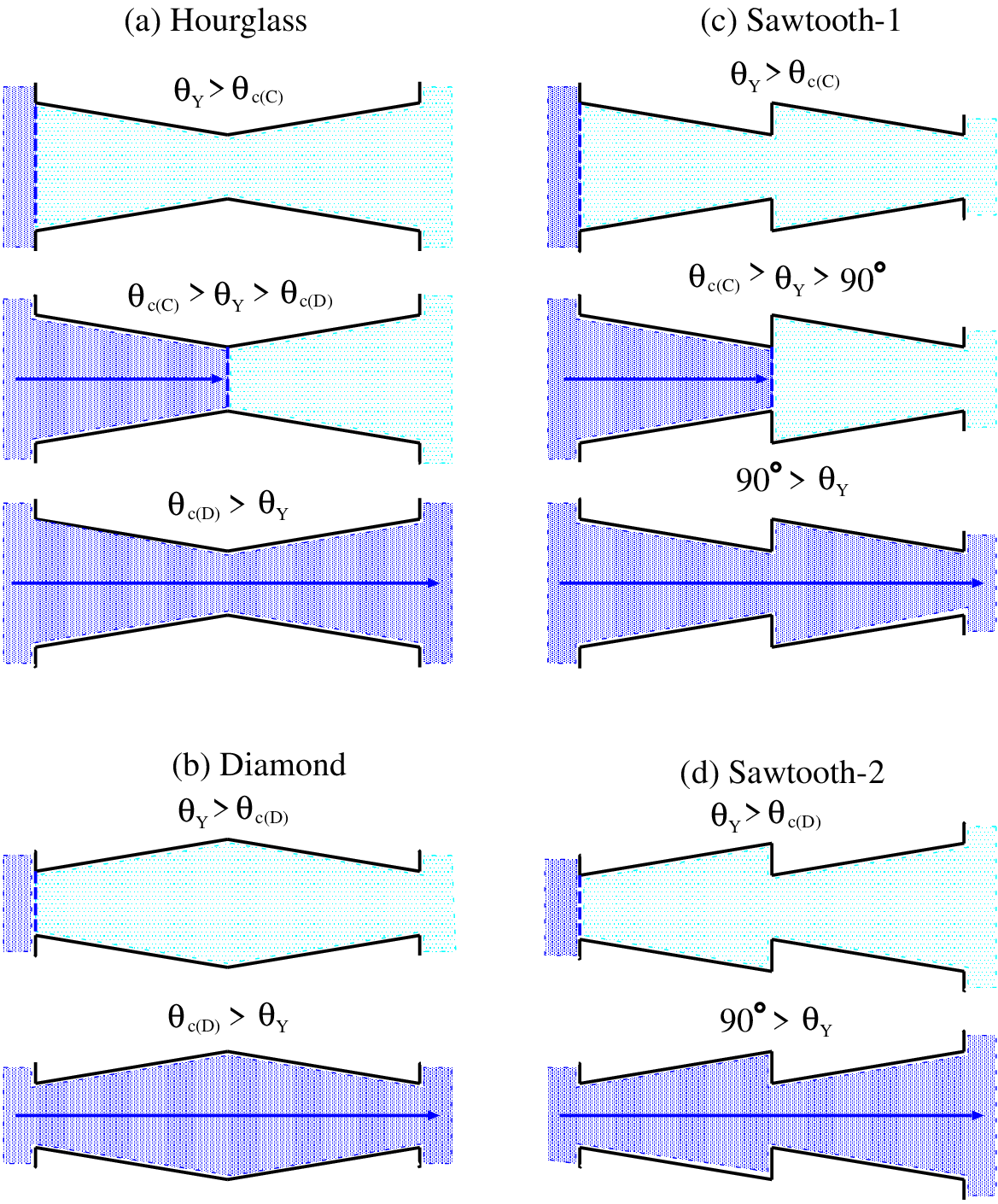}
\end{center}
\caption{
Scenarios of the spontaneous intrusion in double conical capillaries. The dense shadow is liquid and the sparse shadow is vapor. Three scenarios are expected: completely filled, half-filled and completely empty.
} 
\label{fig:D10}
\end{figure}

Figure \ref{fig:D10} summarizes the scenarios of spontaneous intrusion in double conical capillaries. The intrusion occurs from the left liquid reservoir to the right vapor reservoir. Since the spontaneous intrusion can occur only in converging capillaries and not in diverging capillaries when $\theta_{\rm c(C)}>\theta_{\rm Y}>\theta_{\rm c(D)}$, there are three scenarios: complete filling, half-filling, and complete empty. As we consider the quasi-static thermodynamic transient state, the dynamical effects, such as viscous resistance, pinning, vortex etc. at the junction and the entrance are neglected.

Figure~\ref{fig:D10}(a) illustrates the three scenarios in hourglass shaped capillaries. The capillary is completely empty when $\theta_{\rm Y}>\theta_{\rm c(D)}$. It is half-filled when $\theta_{\rm c(C)}>\theta_{\rm Y}>\theta_{\rm c(D)}$, and it is completely filled when $\theta_{\rm c(D)}>\theta_{\rm Y}$. These three scenarios can be confirmed from the free energy landscape in the next subsection. Figure~\ref{fig:D10}(b) illustrates the two scenarios in diamond shaped capillaries: the capillary is completely empty when $\theta_{\rm Y}>\theta_{\rm c(D)}$, and it is completely filled when $\theta_{\rm c(D)}>\theta_{\rm Y}$.

In Figs.~\ref{fig:D10}(c) and (d), we illustrate the scenarios in two sawtooth shaped capillaries. In these cases, a vertical wall at the junction (a shaded pierced-coin shaped vertical wall in Figs.~\ref{fig:D1}(c) and (d)) will affect hydrodynamics of flow. Here, we neglect various hydrodynamic effects and concentrate on the results obtained purely from the thermodynamic free energy consideration.

Figure~\ref{fig:D10}(c) illustrates the three scenarios in sawtooth-1 shaped capillaries (Fig.~\ref{fig:D1}), where the capillary is completely empty when $\theta_{\rm Y}>\theta_{\rm c(C)}$, half-filled when $\theta_{\rm c(C)}>\theta_{\rm Y}>90^{\circ}$, and completely filled when $\theta_{\rm c(D)}>\theta_{\rm Y}$. The half-filling arises because the vertical wall at the junction acts as a free energy barrier or a hydrophobic gate for liquid intrusion when the vertical wall is hydrophobic ($\theta_{\rm Y}>90^{\circ}$), which will be discussed more quantitatively in the next subsection using the free energy landscape. Figure~\ref{fig:D10}(d) illustrates the two scenarios in sawtooth-2 shaped capillary (Fig.~\ref{fig:D1}), where the capillary is completely empty when $\theta_{\rm Y}>\theta_{\rm c(D)}$ and is completely filled when $\theta_{\rm c(D)}>\theta_{\rm Y}$. In contrast to the sawtooth-1 shaped capillary, the vertical wall at the junction does not act as the free energy barrier when $\theta_{\rm c(D)}>\theta_{\rm Y}$ because the wall is hydrophilic as $90^{\circ}>\theta_{\rm c(D)}$. Those scenarios will be the initial state of the forced imbibition which is the subject of the next subsection.

The scenarios illustrated in Figs.~\ref{fig:D10}(c) and (d) are thermodynamic equilibrium states purely judged from the free energy minimum. There are also a completely filled and a half-filled {\it metastable} state, which has the free energy higher than the equilibrium stable states. These metastable states can exist because they are separated from the equilibrium state by the free energy barrier at the junction. This can be clearly seen in the free energy landscape, which is the subject of the next subsection.

\begin{figure}[htbp]
\begin{center}
\includegraphics[width=0.9\linewidth]{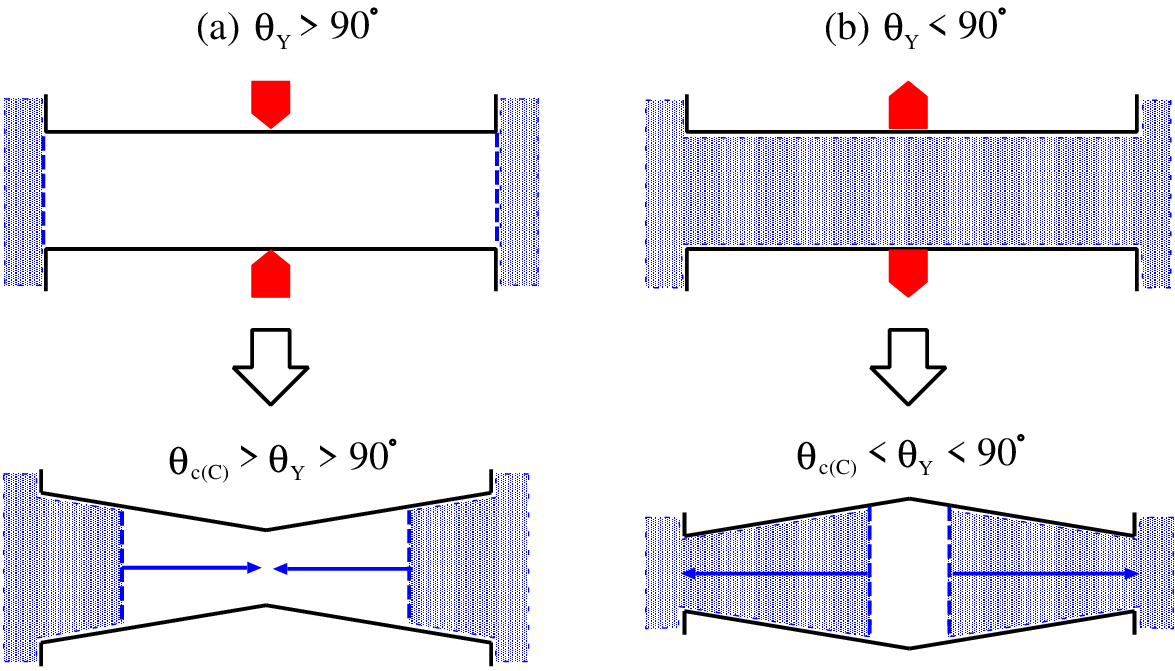}
\end{center}
\caption{
Switching behavior by mechanical deformation of straight cylindrical capillary into (a) converging-diverging hourglass shaped capillary or (b) diverging-converging diamond shaped capillary.
} 
\label{fig:D11}
\end{figure}

If a straight cylindrical capillary could be mechanically deformed~\cite{He2014,Cao2019} into a converging-diverging hourglass or a diverging-converging diamond shaped capillary, we can imagine a mechanical switch illustrated in Fig.~\ref{fig:D11} when both ends (inlet and outlet) of the capillary are immersed in the liquid reservoir. When a hydrophobic straight cylinder is deformed into an hourglass shaped capillary and Young's contact angle satisfies $\theta_{\rm c(C)}>\theta_{\rm Y}>90^{\circ}$, the liquid will intrude from both ends and fill the hourglass shaped capillary (Fig~\ref{fig:D11}(a)). When a hydrophilic straight cylinder completely filled by liquid is deformed into a diamond shaped capillary and Young's contact angle satisfies $\theta_{\rm c(D)}<\theta_{\rm Y}<90^{\circ}$, the liquid will extrude from the capillary and the capillary will be completely empty (Fig~\ref{fig:D11}(b)). Of course, the effect of vapor should be negligible or dissolved into or released from the liquid inside the capillary

\subsection{\label{sec:sec3.2}Forced imbibition in double conical capillaries}

To study the forced imbibition in double conical capillaries, we have to combine the free energy landscape of a single conical capillary considered in section \ref{sec:sec2.3}. The free energy landscape of a double conical capillary is simply a combination of that of a single converging capillary $\tilde{\omega}_{\rm C} (\tilde{z})$ and a diverging capillary $\tilde{\omega}_{\rm D} (\tilde{z})$. We present the free energy landscapes in Figs.~\ref{fig:D12} to \ref{fig:D15}. The parameters $\phi=10^{\circ}$ and $\eta_{\rm C}=4.0$ which characterize the converging and the diverging capillary are the same as those used in section \ref{sec:sec2} so that the critical Young's angles are $\theta_{\rm c(C)}=90+10=100^{\circ}$ and $\theta_{\rm c(D)}=90-10=80^{\circ}$, and the non-dimensional characteristic pressures are given in Tab.~\ref{tab:D2}. The double conical capillaries are twice as long as a single conical capillary considered in section \ref{sec:sec2}. So, we consider the imbibition pathway along $0\leq z\leq 2H$ or $0\leq \tilde{z}\leq 2$ in the non-dimensional unit.  

It is possible to superimpose Fig.~\ref{fig:D9}(a) on \ref{fig:D9}(b) to make a combined phase diagram.  However, it is not straight forward to imagine the imbibition process in double conical capillaries from such a combined phase diagram.  We will continue to use the free energy landscape to discuss the details of the imbibition process.  Most of the symbols and the arrows in Figs.~\ref{fig:D12} to \ref{fig:D15} have the same meanings as those in Fig.~\ref{fig:D8}.

\subsubsection{\label{sec:sec3.2.1}Converging-diverging hourglass shaped capillary}

\begin{figure}[htbp]
\begin{center}
\includegraphics[width=0.9\linewidth]{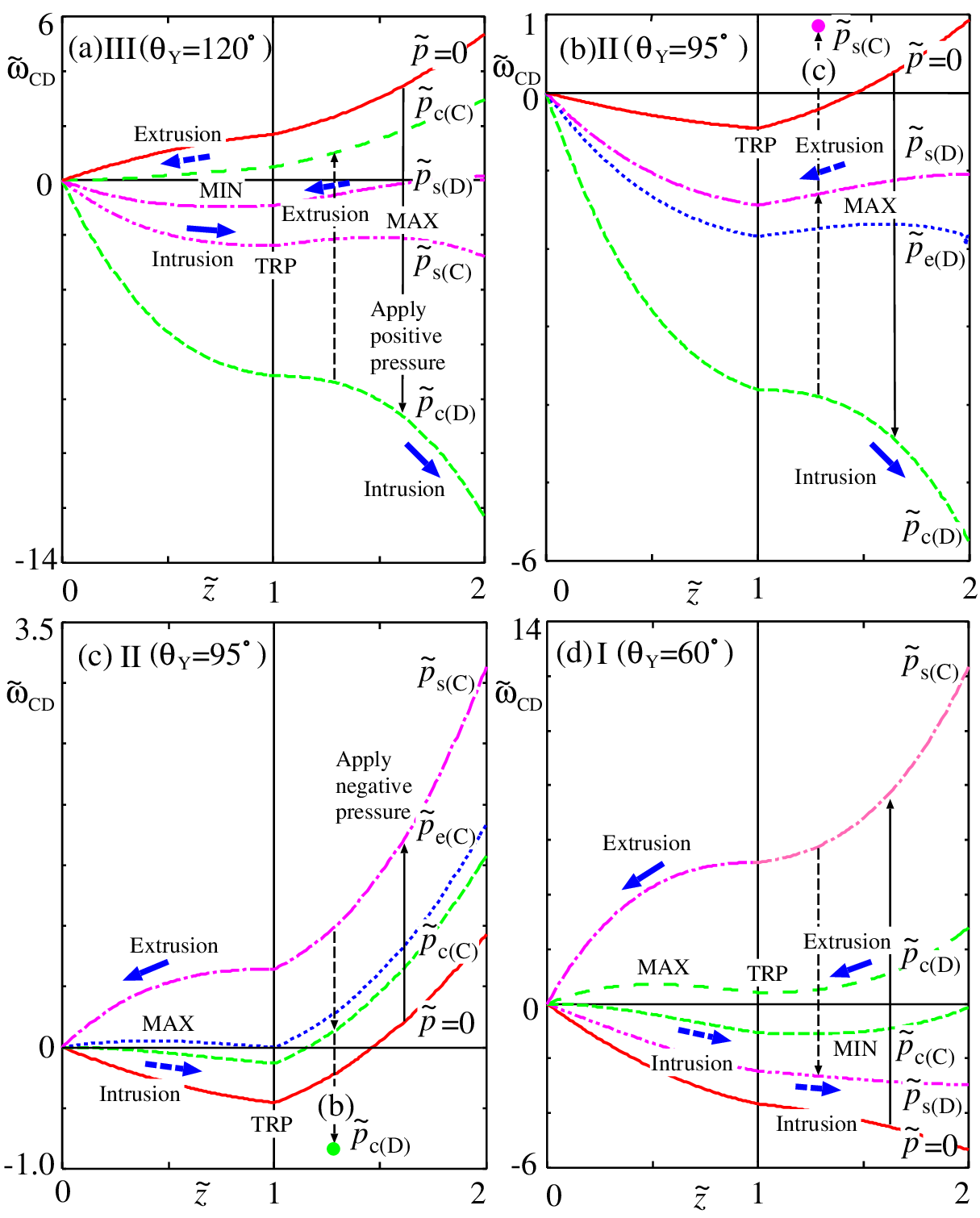}
\end{center}
\caption{
The free energy landscapes of forced imbibition in a converging-diverging hourglass shaped capillary with $\phi=10^{\circ}$ and $\eta_{\rm C}=4.0$ for selected external pressures in Tab.~\ref{tab:D2}. The meanings of the vertical (solid, broken, up and down) arrows in Fig.~\ref{fig:D12} to \ref{fig:D15} are the same as those in Fig.~\ref{fig:D8}.  (a) Intrusion in the region III ($\theta_{\rm Y}=120^{\circ}$). (b) Intrusion in the region II ($\theta_{\rm Y}=95^{\circ}$).  (c) Extrusion in the region II ($\theta_{\rm Y}=95^{\circ}$).  (d) Extrusion in the region I ($\theta_{\rm Y}=60^{\circ}$).
} 
\label{fig:D12}
\end{figure}

The free energy landscapes of a converging-diverging (CD) hourglass shaped capillary (Fig.~\ref{fig:D1}(a)) in the regions I, II and III in Figs.~\ref{fig:D6} and \ref{fig:D7} are presented in Figs.~\ref{fig:D12}(a) to (d), where the non-dimensional free energy $\tilde{\omega}_{\rm CD}$ consists of that of a single converging capillary $\tilde{\omega}_{\rm C}\left(\tilde{z}\right)$ and a diverging capillary $\tilde{\omega}_{\rm D}\left(\tilde{z}\right)$, and is simply given by
\begin{eqnarray}
\tilde{\omega}_{\rm CD}\left(\tilde{z}\right) &=& \tilde{\omega}_{\rm C}\left(\tilde{z}\right),\;\;\; \tilde{z}< 1, \label{eq:D53} \\
&=& \tilde{\omega}_{\rm C}\left(\tilde{z}=1\right)+\tilde{\omega}_{\rm D}\left(\tilde{z}-1\right),\;\;\;
1\le \tilde{z}\le 2.
\label{eq:D54}
\end{eqnarray}
The scenarios of the liquid intrusion and the extrusion in an hourglass shaped capillary predicted from the free energy landscape $\tilde{\omega}_{\rm CD}\left(\tilde{z}\right)$ are summarized as follows (see also Tab.~\ref{tab:D4}).

\renewcommand{\labelenumi}{(\theenumi)}
\begin{enumerate}
\item 
Figure~\ref{fig:D12}(a) presents the free energy landscape of the {\it liquid intrusion} in the region III ($\theta_{\rm Y}>\theta_{\rm c(C)}$) of a converging-diverging hourglass shaped capillary with $\theta_{\rm Y}=120^{\circ}$. Initially ($\tilde{p}=0$) the capillary is completely empty ([E], see Tab.~\ref{tab:D4}) as presented in Fig.~\ref{fig:D10}(a) (see top line with $\tilde{p}=0$). By increasing the (positive) applied pressure from $\tilde{p}=0$, the free energy landscape in the converging part ($0\leq \tilde{z}\leq 1)$) is characterized by a minimum (MIN) and that in the diverging part ($1\leq \tilde{z}\leq 2)$) by a maximum (MAX). Therefore, the intrusion occurs gradually in the converging part until $\tilde{p}_{\rm s(C)}$ is reached. Then the meniscus is trapped by the free energy minimum at the junction (TRP) and the capillary is half-filled ([HF]). By further increasing the pressure, the meniscus jumps from the junction ($\tilde{z}=1$) to the outlet ($\tilde{z}=2$) at $\tilde{p}_{\rm c(D)}$ and the capillary is completely filled ([F]). Therefore, the movement of the liquid-vapor meniscus is {\it gradual} during the first half stage ($0\leq \tilde{p}\leq\tilde{p}_{\rm s(C)}$) and {\it abrupt} in the second half stage ($\tilde{p}_{\rm s(C)}< \tilde{p}\leq\tilde{p}_{\rm c(D)}$) of the process.  The half-filled state of this second half stage becomes thermodynamically metastable before reaching $\tilde{p}_{\rm c(D)}$  In the reverse process of depression, the extrusion occurs {\it abruptly} at $\tilde{p}_{\rm s(D)}$ where the meniscus jumps from the outlet to the inside of converging part so that the capillary is nearly half-filled ([nHF]).  Again, the completely filled state becomes thermodynamically metastable before reaching $\tilde{p}_{\rm s(D)}$.  By further depression, the meniscus moves {\it gradually} from the junction towards the inlet and reaches there at $\tilde{p}_{\rm c(C)}$.  Therefore, we will observe a pressure hysteresis only between $\tilde{p}_{\rm s(D)}$ and $\tilde{p}_{\rm c(D)}$ which is similar to that in a single diverging capillary in section \ref{sec:sec2.3} (cf. Tab.~\ref{tab:D3}(b) and \ref{tab:D4}(a)).

\item 
Figure~\ref{fig:D12}(b) presents the {\it liquid intrusion} into the diverging part of a converging-diverging double conical capillary in the region II ($\theta_{\rm c(C)}>\theta_{\rm Y}>\theta_{\rm c(D)}$) with $\theta_{\rm Y}=95^{\circ}$. Initially, the capillary is half-filled ([HF]) (see middle line of Fig.~\ref{fig:D10}(a)) because the liquid vapor meniscus is trapped by the free energy minimum at the junction (TRP).  When the applied pressure is increased from $\tilde{p}=0$, the meniscus remains trapped at the junction because the free energy maximum (MAX) exists in the diverging part ($1\leq\tilde{z}\leq 2$). The intrusion into the diverging part occurs {\it abruptly} at $\tilde{p}_{\rm c(D)}$ when the MAX disappears. Then the meniscus jumps from the junction to the outlet and the whole capillary is filled by liquid ([F]). In the reverse process of depression, the extrusion occurs {\it abruptly} at $\tilde{p}_{\rm s(D)}$: the meniscus jumps from the outlet to the junction and the capillary becomes half-filled ([HF]).  By further depression (by negative pressure, a vertical arrow with a symbol "(c)" in Fig.~\ref{fig:D12}(b)), the extrusion occurs {\it abruptly} again at $\tilde{p}_{\rm s(C)}$  (<0, see Tab.~\ref{fig:D1}) as presented in Fig.~\ref{fig:D12}(c). We will observe a double pressure hysteresis ranging from the positive $\tilde{p}_{\rm c(D)}$ to the negative $\tilde{p}_{\rm s(C)}$ (Tab.~\ref{tab:D4}(b)).

\item 
Figure~\ref{fig:D12}(c) presents the {\it liquid extrusion} from (vapor intrusion into) the converging part of a converging-diverging double conical capillary in the region II ($\theta_{\rm c(C)}>\theta_{\rm Y}>\theta_{\rm c(D)}$) with $\theta_{\rm Y}=95^{\circ}$. The initial state at $\tilde{p}=0$ is the same ([HF]) as that in Fig.~\ref{fig:D12}(b). The meniscus is trapped by the free energy minimum at the junction (TRP). By increasing the absolute magnitude of the (negative) applied pressure from $\tilde{p}=0$, the complete extrusion ([E]) occurs {\it abruptly} at $\tilde{p}_{\rm s(C)}$ when the free energy maximum (MAX) in the converging part ($0\leq \tilde{z}\leq 1$) disappears. In the reverse process of compression, the intrusion into the converging part occurs {\it abruptly} at $\tilde{p}_{\rm c(C)}$, and the capillary is half-filled ([HF]) again. By further compression (a vertical arrow with a symbol "(b)" in Fig.~\ref{fig:D12}(c)), the complete intrusion occurs {\it abruptly} again at $\tilde{p}_{\rm c(D)}$ (>0, see Tab.~\ref{tab:D2}) as presented in Fig.~\ref{fig:D12}(b). Again, we will observe a double pressure hysteresis ranging from the positive $\tilde{p}_{\rm c(D)}$ to the negative $\tilde{p}_{\rm s(C)}$ (cf. Tab.~\ref{tab:D4}(c) and \ref{tab:D4}(b)).

\item 
Figure~\ref{fig:D12}(d) presents the {\it liquid extrusion} from the whole capillary in the region I ($\theta_{\rm c(D)}>\theta_{\rm Y}$) with $\theta_{\rm Y}=60^{\circ}$. Initially the whole capillary is completely filled by liquid ([F]) at $\tilde{p}=0$ (see bottom line of Fig.~\ref{fig:D10}(a)). The extrusion occurs {\it gradually} by increasing the absolute magnitude of the (negative) pressure in the first half stage due to the existence of minimum (MIN) in the diverging part. The extrusion stops at $\tilde{p}_{\rm c(D)}$ and the meniscus is trapped by the free energy minimum (TRP) at the junction ([HF]). In the second half stage, the extrusion occurs {\it abruptly} at $\tilde{p}_{\rm s(C)}$ when the maximum (Max) in the converging part disappears, and the meniscus jumps from the junction to the inlet ([E]).  In the reverse process of compression, the intrusion starts {\it abruptly} at $\tilde{p}_{\rm c(C)}$ and the meniscus jumps from the inlet to the inside of the diverging part so that the capillary is nearly half-fille ([nHF]). Then, the intrusion proceeds {\it gradually} and the meniscus reaches the outlet at $\tilde{p}_{\rm s(D)}$ ([F]). Similar to Fig.~\ref{fig:D12}(a), we will observe a pressure hysteresis only between $\tilde{p}_{\rm s(C)}$ and $\tilde{p}_{\rm c(C)}$ which is similar but slightly more complex than that in a single converging conical capillary in section \ref{sec:sec2.3} (cf. Tab.~\ref{tab:D3}(d) and \ref{tab:D4}(d)).

\end{enumerate}

\begin{table}
\caption{
Imbibition processes in an hourglass shaped capillary in Fig.~\ref{fig:D12}, which connects the empty [E], the half-filled [HF], the filled [F] state in Fig.~\ref{fig:D10}(a) and the nearly half-filled [nHF] state. The meanings of the other symbols are the same as those in Tab.~\ref{tab:D3}.
}
\label{tab:D4}
\begin{tabular}{cc|cc}\hline\hline
(a)                                  &                                         & (b)                                &  \\
Intrusion                         & Reverse                            &  Intrusion                       & Reverse \\
\hline
                                     &                                        &                                      & $\tilde{p}_{\rm s(C)}$ [E] \\
                                     &                                        &                                      & $\uparrow$a                   \\
$\tilde{p}=0$ [E]              & $\tilde{p}_{\rm c(C)}$ [E]   & $\tilde{p}=0$ [HF]            & $\tilde{p}_{\rm s(D)}$ [HF] \\
$\downarrow$g               & $\uparrow$g                    & $\downarrow$a                & $\uparrow$a  \\
$\tilde{p}_{\rm s(C)}$[HF] & $\tilde{p}_{\rm s(D)}$[nHF]    & $\tilde{p}_{\rm c(D)}$ [F]   & $\tilde{p}_{\rm c(D)}$ [F] \\
$\downarrow$a               & $\uparrow$a                    &                                      & \\
$\tilde{p}_{\rm c(D)}$ [F]  & $\tilde{p}_{\rm c(D)}$ [F]   &  \\
\hline\hline
(c) &                               &  (d)                                 & \\
Extrusion                        & Reverse                          & Extrusion                         &  Reverse \\
\hline
$\tilde{p}_{\rm s(C)}$ [E]  & $\tilde{p}_{\rm s(C)}$ [E]   & $\tilde{p}_{\rm s(C)}$ [E]  & $\tilde{p}_{\rm s(C)}$ [E] \\
$\uparrow$a                  & $\downarrow$a                 & $\uparrow$a                  & $\downarrow$a  \\
$\tilde{p}=0$ [HF]           & $\tilde{p}_{\rm c(C)}$ [HF] & $\tilde{p}_{\rm c(D)}$ [HF]& $\tilde{p}_{\rm c(C)}$ [nHF] \\
                                    & $\downarrow$a                & $\uparrow$g                   & $\downarrow$g \\
                                    & $\tilde{p}_{\rm c(D)}$ [F]   & $\tilde{p}=0$ [F]              & $\tilde{p}_{\rm s(D)}$ [F] \\
\hline\hline
\end{tabular}
\end{table}

\subsubsection{\label{sec:sec3.2.2}Diverging-converging diamond shaped capillary}

\begin{figure}[htbp]
\begin{center}
\includegraphics[width=0.9\linewidth]{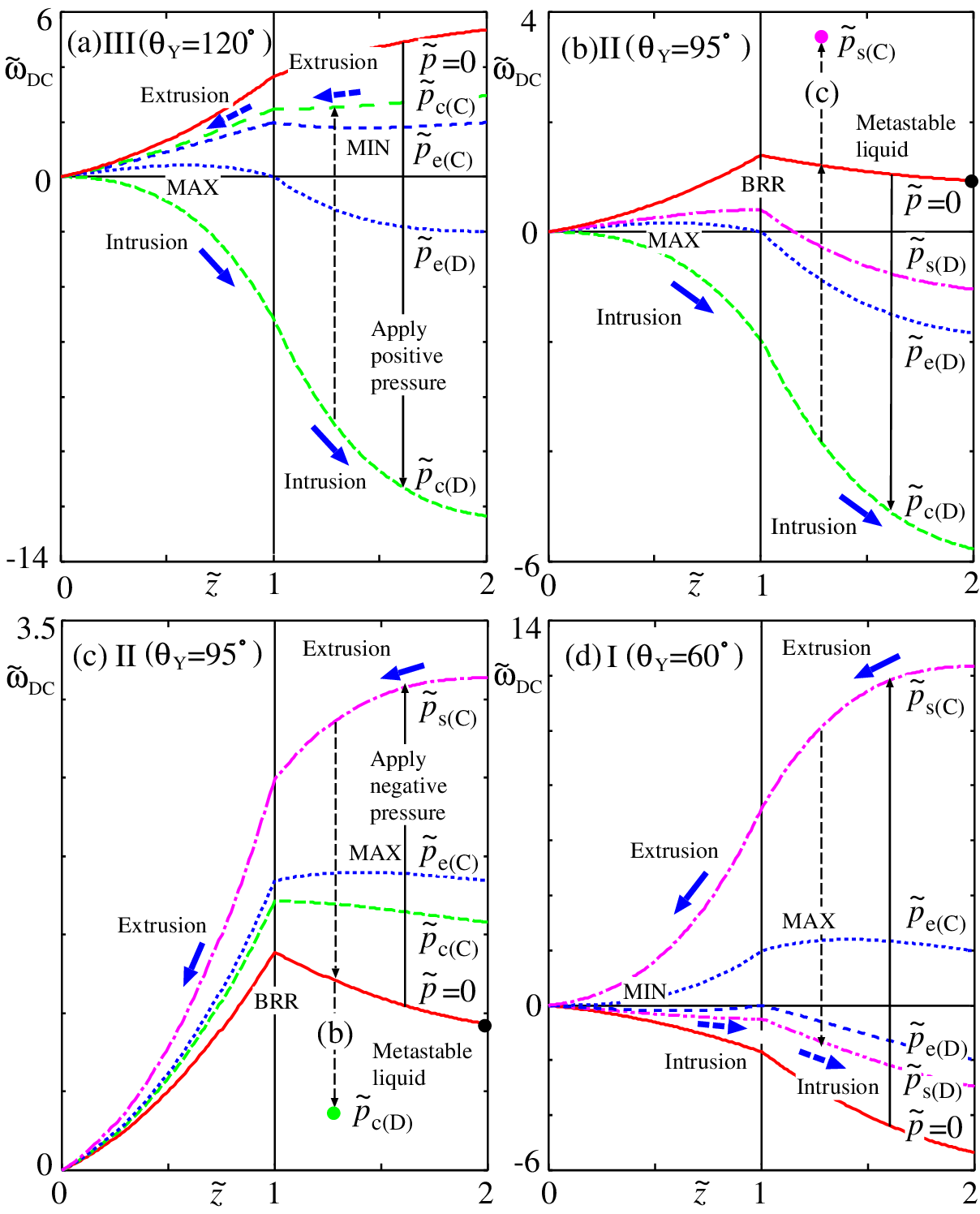}
\end{center}
\caption{
Imbibition in a diverging-converging diamond shaped capillary with $\phi=10^{\circ}$ and $\eta_{\rm C}=4.0$. (a) Intrusion in the region III ($\theta_{\rm Y}=120^{\circ}$). (b) Intrusion in the region II ($\theta_{\rm Y}=95^{\circ}$). (c) Extrusion in the region II ($\theta_{\rm Y}=95^{\circ}$). (d) Extrusion in the region I ($\theta_{\rm Y}=60^{\circ}$).
} 
\label{fig:D13}
\end{figure}

The free energy landscapes of a diverging-converging (DC) diamond shaped capillary (Fig.~\ref{fig:D1}(b)) in the regions I, II and III are presented in Figs.~\ref{fig:D13}(a) to (d), where the free energy $\tilde{\omega}_{\rm DC}$ is given by
\begin{eqnarray}
\tilde{\omega}_{\rm DC}\left(\tilde{z}\right) &=& \tilde{\omega}_{\rm D}\left(\tilde{z}\right),\;\;\; \tilde{z}< 1, \label{eq:D55} \\
&=& \tilde{\omega}_{\rm D}\left(\tilde{z}=1\right)+\tilde{\omega}_{\rm C}\left(\tilde{z}-1\right),\;\;\;
1\le \tilde{z}\le 2. \label{eq:D56}.
\end{eqnarray}
The scenarios of liquid intrusion and extrusion in a diamond shaped capillary predicted from the free energy landscape $\tilde{\omega}_{\rm DC}\left(\tilde{z}\right)$ are summarized in Tab~\ref{tab:D5} and are as follows.

\renewcommand{\labelenumi}{(\theenumi)}
\begin{enumerate}
\item 
Figure~\ref{fig:D13}(a) presents the free energy landscape of the {\it liquid intrusion} in the region III of a diamond shaped capillary with $\theta_{\rm Y}=120^{\circ}$.  The intrusion occurs {\it abruptly} as the free-energy maximum (MAX) exists in the diverging part  ($0\leq \tilde{z}\leq 1$).  The meniscus jumps from the inlet to the outlet at $\tilde{p}_{\rm c(D)}$. In the reverse process of depression, the extrusion in the converging part ($1\leq \tilde{z}\leq 2$) occurs {\it gradually} as the free energy minimum (MIN) exists in the converging part. As soon as the meniscus reaches the junction ($\tilde{z}=1$) at $\tilde{p}_{\rm c(C)}$, the meniscus jumps from the junction to the inlet {\it abruptly} and the extrusion is completed.  Therefore, we will observe a more complex pressure hysteresis than that in an hourglass shaped capillary (cf. Tab.~\ref{tab:D4}(a) and \ref{tab:D5}(a)).

\item 
Figure~\ref{fig:D13}(b) presents the {\it liquid intrusion} in the region II with $\theta_{\rm Y}=95^{\circ}$. Again, the intrusion occurs {\it abruptly} at $\tilde{p}_{\rm c(D)}$ as the free-energy maximum (MAX) appears in the diverging part ($0\leq \tilde{z}\leq 1$). In the reverse process of depression, the meniscus is pinned at the outlet (a black dot in Fig.~\ref{fig:D13}(b)) as the free energy barrier (BRR) exists at the junction, and the whole capillary is filled by the metastable liquid ([MF]) even at $\tilde{p}=0$. By further depression (negative pressure,  a vertical arrow with a symbol "(c)" in Fig.~\ref{fig:D13}(b)), the extrusion is completed ([E]) at $\tilde{p}_{\rm s(C)}$ as presented in Fig.~\ref{fig:D13}(c). Again, we will observe a complex and large pressure hysteresis involving a metastable state (Tab.~\ref{tab:D5}(b)).

\item 
Figure~\ref{fig:D13}(c) presents the {\it liquid extrusion} (vapor intrusion) in the region II with $\theta_{\rm Y}=95^{\circ}$. The whole capillary is empty in the thermodynamic equilibrium. However, the meniscus could be pinned at the outlet (a black dot in Fig.~\ref{fig:D13}(c)) e.g., by the reverse process in (b) and the whole capillary could be filled by the metastable liquid ([MF]) as the free energy barrier (BRR) exists at the junction. Extrusion of this metastable liquid occurs {\it abruptly} at $\tilde{p}_{\rm s(C)}$ as the free energy maximum (MAX) exists in the converging part. The reverse process of the intrusion occurs {\it abruptly} not at $\tilde{p}=0$ but at higher pressure $\tilde{p}_{\rm c(D)}$ (a vertical arrow with a symbol "(b)" in Fig.~\ref{fig:D13}(c)) as presented in Fig.~\ref{fig:D13}(b). We will observe a complex and large pressure hysteresis similar to that in Fig.~\ref{fig:D13}(b) (Tab.~\ref{tab:D5}(c)).

\item 
Figure~\ref{fig:D13}(d) presents the {\it liquid extrusion} in the region I with $\theta_{\rm Y}=60^{\circ}$. Initially, the thermodynamically stable liquid occupies the whole capillary. Extrusion occurs {\it abruptly} at $\tilde{p}_{\rm s(C)}$ as there exists the free energy maximum (MAX) in the converging part. In the reverse process of compression, the intrusion into the diverging part occurs {\it gradually} as there exist free energy minimum (MIN) in the diverging part. At $\tilde{p}_{\rm s(D)}$ the meniscus reaches the junction. Then, it {\it abruptly} jumps to the outlet. Therefore, we will observe a complex pressure hysteresis similar to that in Fig.~\ref{fig:D13}(a) (Tab.~\ref{tab:D5}(d)).

\end{enumerate}

\begin{table}
\caption{
Imbibition processes in a diamond shaped capillary in Fig.~\ref{fig:D13}, which connects the empty [E], the metastable filled [MF], and the filled [F] state. Letters "a", "g", and "g+a" beside the arrows indicate that the imbibition occurs {\it abruptly} (a), {\it gradually} (g), and {\it gradually} then {\it abruptly} (g+a).
}
\label{tab:D5}
\begin{tabular}{cc|cc}\hline\hline
(a)                                 &                                      & (b)                                &  \\
Intrusion                        & Reverse                          &  Intrusion                      &  Reverse \\
\hline
                                    &                                       &                                     & $\tilde{p}_{\rm s(C)}$ [E] \\
                                    &                                       &                                     & $\uparrow$a   \\
$\tilde{p}=0$ [E]             &  $\tilde{p}_{\rm c(C)}$ [E]  & $\tilde{p}=0$ [E]             & $\tilde{p}=0$ [MF] \\
$\downarrow$a              & $\uparrow$g+a                    & $\downarrow$a               & $\uparrow$ \\
$\tilde{p}_{\rm c(D)}$ [F]  & $\tilde{p}_{\rm c(D)}$ [F]   & $\tilde{p}_{\rm c(D)}$ [F] & $\tilde{p}_{\rm c(D)}$ [F]  \\
\hline\hline
(c)                                &                                       & (d)                                  & \\
Extrusion                      & Reverse                           & Extrusion                        & Reverse \\
\hline
$\tilde{p}_{\rm s(C)}$ [E] &  $\tilde{p}_{\rm s(C)}$ [E]  & $\tilde{p}_{\rm s(C)}$ [E]   & $\tilde{p}_{\rm s(C)}$ [E]\\
$\uparrow$a                  & $\downarrow$                 & $\uparrow$a                    & $\downarrow$g+a  \\
$\tilde{p}=0$ [MF]         & $\tilde{p}=0$ [E]                & $\tilde{p}=0$ [E]              & $\tilde{p}_{\rm s(D)}$ [F]  \\
                                   & $\downarrow$a                 &                                      &                                      \\
                                   & $\tilde{p}_{\rm c(D)}$ [F]    &                                      &                                      \\
\hline\hline
\end{tabular}
\end{table}

Therefore, a subtle difference in the shape of the hourglass shaped and the diamond shaped capillary leads to a dramatic change in intrusion and extrusion behaviors.

\subsubsection{Converging-converging sawtooth-1 shaped capillary}

\begin{figure}[htbp]
\begin{center}
\includegraphics[width=0.9\linewidth]{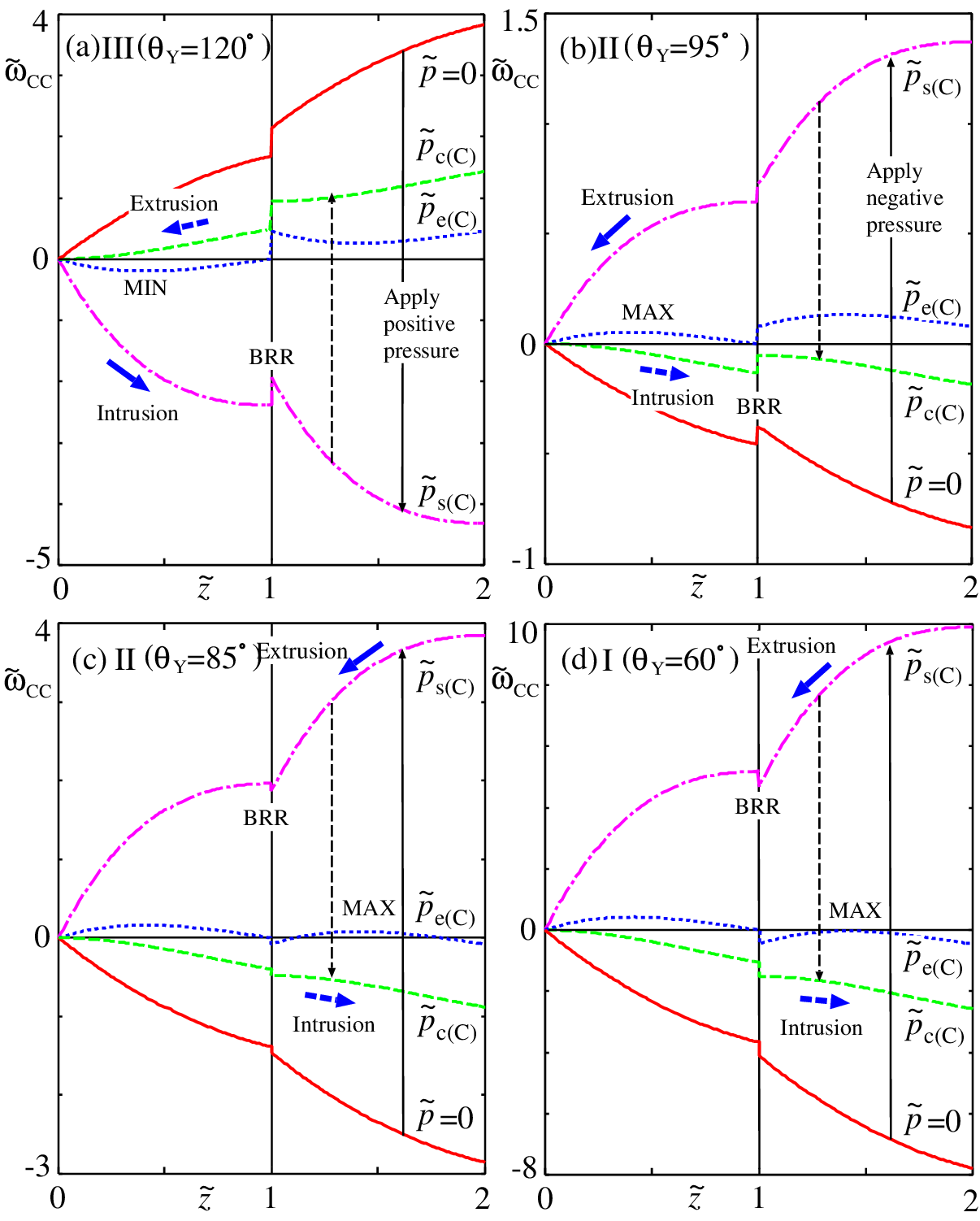}
\end{center}
\caption{
Imbibition in a converging-converging swatooth-1 shaped capillary with $\phi=10^{\circ}$ and $\eta_{\rm C}=4.0$.    (a) intrusion in the region III ($\theta_{\rm Y}=120^{\circ}$), (b) extrusion in the region II when the pierced-coin shaped vertical wall at the junction is hydrophobic ($\theta_{\rm Y}=95^{\circ}$), (c) extrusion in the region II when the vertical wall is hydrophilic ($\theta_{\rm Y}=85^{\circ}$), and (d) extrusion in the region I ($\theta_{\rm Y}=60^{\circ}$).
} 
\label{fig:D14}
\end{figure}

The free energy landscapes of a converging-converging (CC) sawtooth-1 shaped capillary (Fig.~\ref{fig:D1}(c)) in the regions I, II and III are presented in Figs.~\ref{fig:D14}(a) to (d), where the free-energy $\tilde{\omega}_{\rm CC}$ is given by
\begin{eqnarray}
\tilde{\omega}_{\rm CC}\left(\tilde{z}\right) &=& \tilde{\omega}_{\rm C}\left(\tilde{z}\right),\;\;\; \tilde{z}< 1, \label{eq:D57} \\
&=& \Delta \tilde{\omega}_{\rm wl} + \tilde{\omega}_{\rm C}\left(\tilde{z}=1\right)+\tilde{\omega}_{\rm C}\left(\tilde{z}-1\right),
1\le \tilde{z}\le 2, \label{eq:D58}
\end{eqnarray}
and we have added an extra wall-liquid interaction energy written in original unit as
\begin{equation}
\Delta \omega_{\rm wl}=-\pi \left(R_{\rm C}^{2}(0)-R_{\rm D}^{2}(0)\right)\gamma_{\rm lv}\cos\theta_{\rm Y},
\label{eq:D59}
\end{equation}
which originates from the pierced-coin shaped vertical wall (the shaded wall at the junction in Figs.~\ref{fig:D1}(c) and (d)) by appropriate scaling in Eq.~(\ref{eq:D27}). This wall acts as an up step with $\Delta \omega_{\rm wl}>0$ if the wall is hydrophobic ($\theta_{\rm Y}>90^{\circ}$) or a down step with $\Delta \omega_{\rm wl}<0$ if the wall is hydrophilic ($\theta_{\rm Y}<90^{\circ}$). They may act as a barrier (BRR) for imbibition.  Note that this barrier is not due to the hydrophobic barrier by the heterogeneous nucleation of bubble~\cite{Tinti2017,Tinti2023,Remsing2015}, but rather due to the simple free-energy or potential barrier~\cite{Trick2014}. The scenarios of liquid intrusion and extrusion in a sawtooth-1 shaped capillary are summarized in Tab~\ref{tab:D6} and are as follows.

\renewcommand{\labelenumi}{(\theenumi)}
\begin{enumerate}
\item 
Figure~\ref{fig:D14}(a) presents the {\it liquid intrusion} in the region III of a converging-converging sawtooth-1 shaped capillary with $\theta_{\rm Y}=120^{\circ}$.  The intrusion in the first converging part ($0\leq \tilde{z}\leq 1$) occurs {\it gradually} as the free energy minimum (MIN) exists.  Then, the intrusion stops at the junction ($\tilde{z}=1$) and the capillary is half-filled ([HF]) at $\tilde{p}_{\rm s(C)}$ because of the free energy barrier (BRR) $\Delta \omega_{\rm wl}>0$ (Eq.~(\ref{eq:D59})) by the hydrophobic vertical wall.  In the reverse process, the extrusion from the first converging part occurs {\it gradually} and is completed at $\tilde{p}_{\rm c(C)}$ (Tab.~\ref{tab:D6}(a)).  Of course, a small perturbation such as mechanical vibration, thermal fluctuation etc. would help liquid to overcome the barrier and to spill out from the hole of the vertical wall.  Then, the whole capillary would be filled by liquid.

\item 
Figure~\ref{fig:D14}(b) presents the {\it liquid extrusion} (vapor intrusion) in the region II with $\theta_{\rm Y}=95^{\circ}$. Initially, only the first converging part is filled by liquid ([HF]) (see the middle line of Fig.~\ref{fig:D10}(c)). The intrusion into the second converging part ($1\leq \tilde{z}\leq 2$) is prohibited due to the free-energy barrier (BRR) $\Delta \omega_{\rm wl}>0$ at the junction because $\theta_{\rm c(C)}>\theta_{\rm Y}>90^{\circ}$. By increasing the magnitude of the negative applied pressure, the liquid extrusion from the first part occurs {\it abruptly} at $\tilde{p}_{\rm s(C)}$ because of the free energy maximum (MAX), which disappears at $\tilde{p}_{\rm s(C)}$. Even if the second part is also filled by liquid, the extrusion from whole capillary occurs {\it abruptly} also at $\tilde{p}_{\rm s(C)}$ because the barrier at the junction is simply a descending step. In the reverse process of compression, the intrusion into the first part occurs abruptly at $\tilde{p}_{\rm c(C)}$. Therefore, we will observe a pressure hysteresis which is similar to that in a single converging capillary in section \ref{sec:sec2.3} (cf. Tab.~\ref{tab:D3}(c) and \ref{tab:D6}(b)).

\item 
Figure~\ref{fig:D14}(c) presents the {\it liquid extrusion} in the region II with $\theta_{\rm Y}=85^{\circ}$. Initially the whole capillary is filled ([F]) by liquid (see the bottom line of Fig.~\ref{fig:D10}(c)) because the energy $\Delta \omega_{\rm wl}<0$ is a descending step for $90^{\circ}>\theta_{\rm Y}$. By increasing the magnitude of the negative applied pressure, the liquid extrusion in the second part ($1\leq \tilde{z}\leq 2$) occurs {\it abruptly} at $\tilde{p}_{\rm s(C)}$ because the free energy maximum (MAX) exists. However, the meniscus is pinned at the junction by the energy barrier (BRR) $\Delta \omega_{\rm wl}<0$, and the sawtooth-1 shaped capillary will be half-filled ([HF]) by metastable liquid. If the meniscus can be freed from the BRR by some perturbations, the extrusion from the second part would be followed by the extrusion from the first part.  In the reverse process of compression, the intrusion into the second part occurs abruptly at $\tilde{p}_{\rm c(C)}$. Again, we will observe a pressure hysteresis which is similar to that in a single converging capillary in section \ref{sec:sec2.3} (cf. Tab.~\ref{tab:D3}(c) and \ref{tab:D6}(c)).

\item 
Figure~\ref{fig:D14}(d) presents the {\it liquid extrusion} in the region I with $\theta_{\rm Y}=60^{\circ}$. Initially, the whole capillary is filled by liquid ([F]) (see the middle line of Fig.~\ref{fig:D10}(c)). The extrusion in the second part occurs {\it abruptly} at $\tilde{p}_{\rm s(C)}$ as there exist the free-energy maximum (MAX). However, again, the meniscus is pinned at the junction due to the energy barrier (BRR) $\Delta \omega_{\rm wl}<0$ and the capillary is half-filled ([HF]) by metastable liquid. In the reverse process of compression, the intrusion into the second part occurs {\it abruptly} at $\tilde{p}_{\rm c(C)}$ as the energy $\Delta \omega_{\rm wl}<0$ acts as a descending step. Again, we will observe a pressure hysteresis (cf. Tab.~\ref{tab:D3}(c) and ~\ref{tab:D6}(d)).

\end{enumerate}

Therefore, ratchet-like character appears in converging-converging sawtooth-1 shaped capillaries: full extrusion of liquid from the whole capillary (empty state) is possible but the reverse process of full intrusion (filled state) is not possible in Figs.~\ref{fig:D14}(a) and (b) (Tabs.~\ref{tab:D6}(a) and (b)), while full extrusion (empty state) is not possible but the reverse process of full intrusion (filled state) is possible in Figs.~\ref{fig:D14}(c) and (d) (Tabs.~\ref{tab:D6}(c) and (d)).

\begin{table}
\caption{
Imbibition processes in a sawtooth-1 shaped capillary in Fig.~\ref{fig:D14}, which connects the empty [E], the half-filled [HF], and the filled [F] state in Fig.~\ref{fig:D10}(c). 
}
\label{tab:D6}
\begin{tabular}{cc|cc}\hline\hline
(a)                                &                                      & (b)                                   &  \\
Intrusion                       & Reverse                          & Extrusion                         & Reverse \\
\hline
$\tilde{p}=0$ [E]            & $\tilde{p}_{\rm c(C)}$ [E]  & $\tilde{p}_{\rm s(C)}$ [E]    &   $\tilde{p}_{\rm s(C)}$ [E] \\
$\downarrow$g              & $\uparrow$g                   & $\uparrow$a       & $\downarrow$a \\
$\tilde{p}_{\rm s(C)}$ [HF] & $\tilde{p}_{\rm s(C)}$ [HF]  & $\tilde{p}=0$ [HF]  & $\tilde{p}_{\rm c(C)}$ [HF] \\
\hline\hline
(c) &  & (d) & \\
Extrusion                        & Reverse                           & Extrusion             & Reverse \\
\hline
$\tilde{p}_{\rm s(C)}$ [HF] & $\tilde{p}_{\rm s(C)}$ [HF] & $\tilde{p}_{\rm s(C)}$ [HF] & $\tilde{p}_{\rm s(C)}$ [HF]  \\
$\uparrow$a                    &  $\downarrow$a                & $\uparrow$a         & $\downarrow$a  \\
$\tilde{p}=0$ [F]              & $\tilde{p}_{\rm c(C)}$ [F]         & $\tilde{p}=0$ [F]    & $\tilde{p}_{\rm c(C)}$ [F] \\
\hline\hline
\end{tabular}
\end{table}

\subsubsection{Diverging-diverging sawtooth-2 shaped capillary}

\begin{figure}[htbp]
\begin{center}
\includegraphics[width=0.9\linewidth]{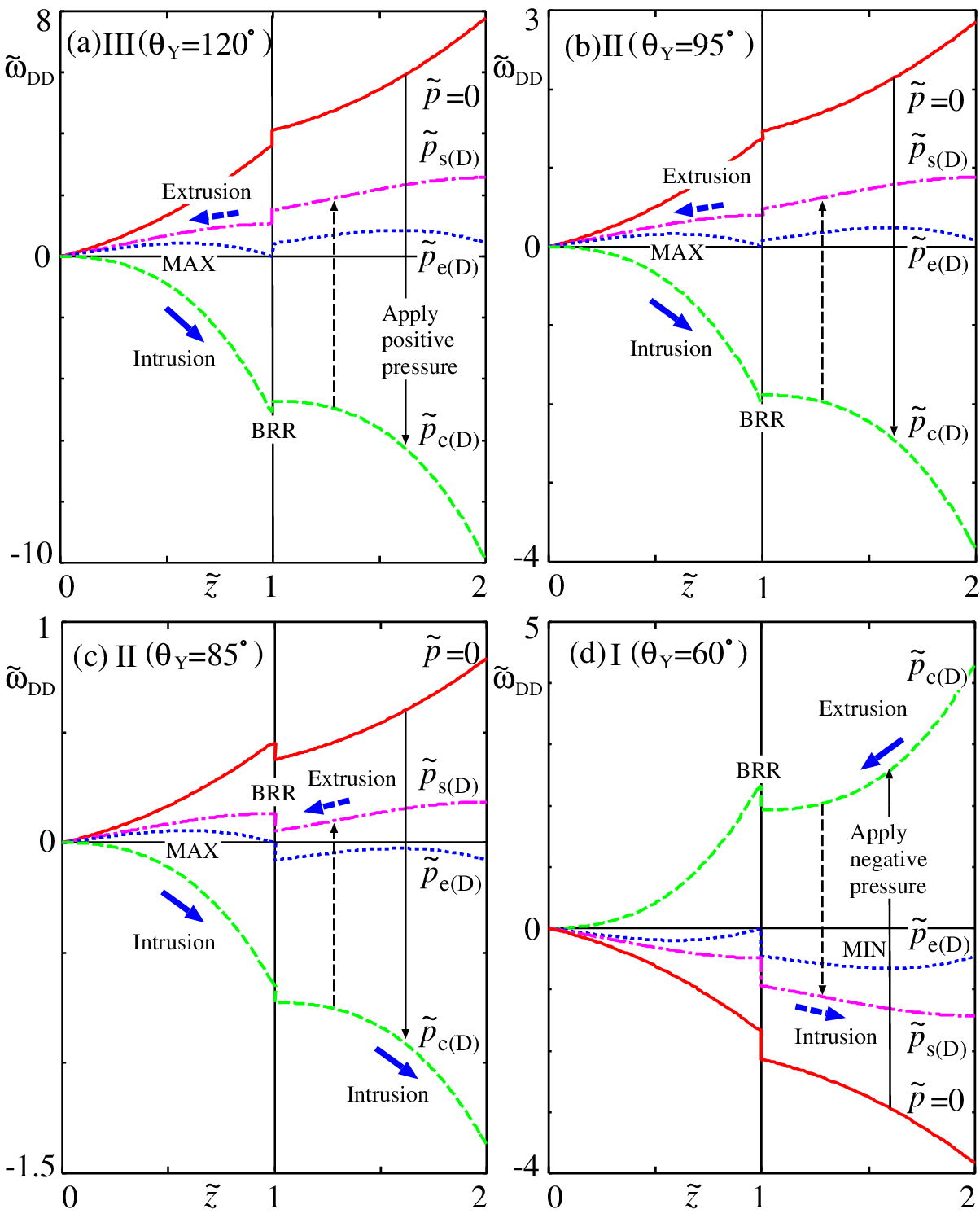}
\end{center}
\caption{
Imbibition in a swatooth-2 shaped capillary with $\phi=10^{\circ}$ and $\eta_{\rm C}=4.0$.   (a) intrusion in the region III ($\theta_{\rm Y}=120^{\circ}$), (b) extrusion in the region II when the vertical wall at the junction is hydrophobic ($\theta_{\rm Y}=95^{\circ}$), (c)  extrusion in the region II when the vertical wall is hydrophilic ($\theta_{\rm Y}=85^{\circ}$), and (d) extrusion in the region I ($\theta_{\rm Y}=60^{\circ}$).
} 
\label{fig:D15}
\end{figure}

The free energy landscapes of a diverging-diverging (DD) sawtooth-2 capillary (Fig.~\ref{fig:D1}(d)) in the regions I, II and III are presented in Figs.~\ref{fig:D15}(a) to (d), where the free-energy $\tilde{\omega}_{\rm DD}$ is given by
\begin{eqnarray}
\tilde{\omega}_{\rm DD}\left(\tilde{z}\right) &=& \tilde{\omega}_{\rm D}\left(\tilde{z}\right),\;\;\; \tilde{z}< 1, \label{eq:D60} \\
&=& \Delta \tilde{\omega}_{\rm wl} + \tilde{\omega}_{\rm D}\left(\tilde{z}=1\right)+\tilde{\omega}_{\rm D}\left(\tilde{z}-1\right),
1\le  \tilde{z}\le 2, \label{eq:D61}
\end{eqnarray}
where $\Delta \tilde{\omega}_{\rm wl}$ is the contribution form the vertical wall at the junction given by Eq.~(\ref{eq:D59}).

\renewcommand{\labelenumi}{(\theenumi)}
\begin{enumerate}
\item 
Figure~\ref{fig:D15}(a) presents the free energy landscape of the {\it liquid intrusion} in the region III of a diverging-diverging sawtooth-2 shaped capillary with $\theta_{\rm Y}=120^{\circ}$. The intrusion in the first diverging part  ($0\leq \tilde{z}\leq 1$) occurs {\it abruptly} at $\tilde{p}_{\rm c(D)}$ as the free-energy maximum (MAX) exists in the first part. Then, the intrusion in the second diverging part ($1\leq \tilde{z}\leq 2$) could follow. However, there exists the free energy barrier (BRR) $\Delta \omega_{\rm wl}>0$, and the intrusion into the second diverging part is prohibited, and the capillary is half-filled ([HF]). In the reverse process of depression, the extrusion from the first part occurs abruptly at $\tilde{p}_{\rm s(D)}$. Therefore, we will observe a pressure hysteresis which is similar to that in a single diverging capillary in section \ref{sec:sec2.3} (cf. Tab.~\ref{tab:D3}(b) and \ref{tab:D6}(a)). Again, a small perturbation would help the liquid to spill out from the junction (pierce of coin).

\item 
Figure~\ref{fig:D15}(b) presents the {\it liquid intrusion} in the region II with $\theta_{\rm Y}=95^{\circ}$. The landscape is very similar to that in Fig.~\ref{fig:D15}(a) except for a smaller energy barrier $\Delta \omega_{\rm wl}>0$ at the junction; the intrusion and the extrusion is almost the same as that in (a) (Tab.~\ref{tab:D6}(b)).

\item 
Figure~\ref{fig:D15}(c) presents the {\it liquid intrusion} in the region II with $\theta_{\rm Y}=85^{\circ}$. Initially, the whole capillary is empty ([E]) as illustrated in Fig.~\ref{fig:D10}(d) for $\theta_{\rm c(D)}<\theta_{\rm Y}$. By increasing the magnitude of the positive applied pressure, the liquid intrusion into the whole capillary ([F]) occurs {\it abruptly} at $\tilde{p}_{\rm c(D)}$ as the free energy maximum (MAX) exist. Now, the energy step $\Delta \omega_{\rm wl}<0$ at the junction plays no role. In the reverse process of depression, the extrusion from the second part ($1\leq \tilde{z}\leq 2$) occurs abruptly at $\tilde{p}_{\rm s(D)}$, but the meniscus is pinned by the barrier (BRR) at the junction. The capillary is half-filled ([HF]) and the first part is filled by metastable liquid. Therefore, we will observe a complex pressure hysteresis which involves the completely empty ([E]), the completely filled ([F]), and the metastable half-filled ([HF]) state (Tab.~\ref{tab:D6}(c)).

\item 
Figure~\ref{fig:D15}(d) presents the {\it liquid extrusion} in the region I with $\theta_{\rm Y}=60^{\circ}$. Initially, the thermodynamically stable liquid occupies the whole capillary ([F]). The extrusion in the second part ($1\leq \tilde{z}\leq 2$) occurs {\it gradually} as there exists the free-energy minimum (MIN). The extrusion in the second part is completed at $\tilde{p}_{\rm c(D)}$ However, the subsequent extrusion from the first part ($0\leq \tilde{z}\leq 1$) cannot occur as the energy $\Delta \omega_{\rm wl}<0$ acts as the barrier (BRR) and the meniscus is pinned at the junction. The first converging part is filled by metastable liquid ([HF]). In the reverse process of compression, the intrusion into the second part occurs {\it gradually} at $\tilde{p}_{\rm s(D)}$. We will observe a complex pressure hysteresis (Tab.~\ref{tab:D6}(d)).
 
\end{enumerate}

\begin{table}
\caption{
Imbibition process in a sawtooth-2 shaped capillary in Fig.~\ref{fig:D15}, which connects the empty [E], the half-filled [HF], and the filled [F] state in Fig.~\ref{fig:D10}(d). 
}
\label{tab:D7}
\begin{tabular}{cc|cc}\hline\hline
(a)                                    &                                         & (b)  & \\
Intrusion                           & Reverse                             &  Intrusion                       & Reverse \\
\hline
$\tilde{p}=0$ [E]                 & $\tilde{p}_{\rm s(D)}$ [E]    & $\tilde{p}=0$ [E]              & $\tilde{p}_{\rm s(D)}$ [E] \\
$\downarrow$a                 & $\uparrow$a                      & $\downarrow$a                 & $\uparrow$a \\
$\tilde{p}_{\rm c(D)}$ [HF]   & $\tilde{p}_{\rm c(D)}$ [HF]   & $\tilde{p}_{\rm c(D)}$ [HF] & $\tilde{p}_{\rm c(D)}$ [HF] \\
\hline\hline
(c)                                    &                                         & (d)  &\\
Intrusion                           & Reverse                             & Extrusion                         & Reverse \\
\hline
$\tilde{p}=0$ [E]               & $\tilde{p}_{\rm s(D)}$ [HF]    & $\tilde{p}_{\rm c(D)}$ [HF]  & $\tilde{p}_{\rm c(D)}$ [HF]  \\
$\downarrow$a                 &  $\uparrow$a                      & $\uparrow$g                     & $\downarrow$g  \\
$\tilde{p}_{\rm c(D)}$ [F]   & $\tilde{p}_{\rm c(D)}$ [F]      & $\tilde{p}=0$ [F]                & $\tilde{p}_{\rm s(D)}$ [F] \\
\hline\hline
\end{tabular}
\end{table}

Again, the ratchet-like character appears in a diverging-diverging sawtooth-2 shaped capillary: the full extrusion of liquid from the whole capillary (empty state) is possible but the reverse process of full intrusion (filled state) is not possible in Figs.~\ref{fig:D15}(a) and (b) (Tabs.~\ref{tab:D7}(a) and (b)), while the full extrusion (empty state) is not possible but the reverse full intrusion (filled state) is possible in Figs.~\ref{fig:D14}(c) and (d) (Tabs.~\ref{tab:D7}(c) and (d)).  In these two sawtooth shaped capillaries (sawtooth-1 and sawtooth-2), the pierced-coin shaped vertical wall at the junction could acts as a barrier in the free energy landscape because the wall would adsorb (hydrophilic) or repel (hydrophobic) liquid. Although we concentrated on the (static) thermodynamics and considered the imbibition process from the free energy landscape, this vertical wall will play very complex role in hydrodynamics.

So far, we have concentrated on the thermodynamic aspect of imbibition in conical and double-conical capillaries, and have not considered the kinetic and the hydrodynamic aspect.  Once the imbibition starts and the steady flow is established, we can consider the hydrodynamics. Our thermodynamic results predict the condition of the onset of the spontaneous and the forced imbibition.  Therefore, our thermodynamic results would be the starting point for designing new experimental and numerical studies of hydrodynamics in realistic systems with structures similar to double conical structures even though there have already been some reports~\cite{Gravelle2013,Balannec2018,Antunes2022,Trick2014,Li2019,He2014,Cao2019,Goli2022,Goli2019}.

To study the hydrodynamics in double conical capillaries, a theoretical approach assuming the fully-developed laminar flow following Hagen-Poiseulle law, which has been used to study the steady flow in conical capillaries~\cite{Reyssat2008,Urteaga2013,Berli2014,Gorce2016,Singh2020,Wu2021,Iwamatsu2022}, would be possible.  However, such an approach might not be reliable for double conical capillaries because the steady laminar flow may not be established~\cite{Antunes2022} because of the existence of the junction of two conical capillaries.  Furthermore, even in cylindrical and conical capillaries, the standard non-slip boundary condition~\cite{Landau1987} may not be applicable~\cite{Gravelle2013,Tran-Duc2019,Zhang2020,Mondal2021} when the capillary radius becomes nanoscale.  Furthermore, the dissipation at the inlet (entrance) or outlet (exit)~\cite{Sampson1891,Weissberg1962,Suk2017,Heiranian2020} could not be negligible.  Nevertheless, our thermodynamic results would be the basis for the hydrodynamics studies of double conical capillaries.

\section{\label{sec:sec4} Conclusion}

In this study, we considered the thermodynamics of spontaneous as well as forced imbibition of liquid in capillaries of double conical structures with hourglass, diamond, and sawtooth shapes, which are the prototypes of various natural as well as artificial micro and nano-fluidic systems. We found that the spontaneous intrusion of liquid can occur when Young's contact angle is smaller than the critical Young's contact angle determined from the modified Laplace pressure. The critical contact angle for the onset of spontaneous imbibition of the converging and the diverging capillary belong to the hydrophobic and the hydrophilic region, respectively, and they are determined from the tilt angle of the capillary wall. This asymmetry between the converging and the diverging capillary gives functionality not only to the single conical capillaries~\cite{Singh2020,Iwamatsu2022} but to the double conical capillaries.

The free energy landscape of forced imbibition is studied by assuming the imbibition pathway with a constant Young's contact angle. Even though the condition of the onset of forced imbibition is simply given by the condition that the applied pressure overcomes the highest Laplace pressure at the inlet or the outlet where the capillary is narrowest, the free energy landscape is complex and exhibits either a maximum or a minimum, which suggests either an abrupt imbibition with a pressure hysteresis or a gradual and continuous imbibition. Furthermore, because of the four combinations of the converging and the diverging capillary of the double conical structures, various scenarios of the liquid intrusion and the liquid extrusion including the appearance of a metastable filled and a half-filled state are suggested from the free energy landscapes. These findings would be beneficial in elucidating various imbibition processes in nature and developing functioning micro- and nano-capillaries of artificial double conical structures.

\section*{Author Declaration}
\subsection*{Conflict of interest}
The author declares no conflict of interest.

\section*{Data Availability Statement}
The data that support the findings of this study are available from the author upon reasonable request.

\bibliography{POF23V3}

\end{document}